\documentclass[useAMS,usenatbib]{mn2e}
\usepackage{graphicx}

\def\gsim{\;\lower.6ex\hbox{$\sim$}\kern-6.7pt\raise.4ex\hbox{$>$}\;}
\def\lsim{\;\lower.6ex\hbox{$\sim$}\kern-6.7pt\raise.4ex\hbox{$<$}\;}

\title[Cepheid pulsation models in WFC3 filters]{Cepheid theoretical models and observations in HST/WFC3 filters: the effect on the Hubble constant H$_0$}

\author[G. Fiorentino et al.]{Giuliana Fiorentino$^{1,2}$\thanks{E-mail:giuliana.fiorentino@oabo.inaf.it}, Ilaria Musella$^{3}$ and Marcella Marconi$^{3}$\\
$^{1}$INAF-Osservatorio Astronomico di Bologna, via Ranzani 1, 40127, Bologna, Italy\\
$^{2}$Dipartimento di Astronomia, Universit\`{a} di Bologna, via Ranzani 1, 40127, Bologna,  Italy\\
$^{3}$INAF, Osservatorio Astronomico di Capodimonte,via Moiariello 16, 80131, Napoli, Italy}

\begin{document}

\date{Accepted XX. Received XX; in original form 2013 April 29}

\pagerange{\pageref{firstpage}--\pageref{lastpage}} \pubyear{2013}

\maketitle

\label{firstpage}

\begin{abstract}
We present a complete theoretical scenario for classical Cepheids in the
most commonly used HST/WFC3 filters, going from optical (F555W, F606W and F814W) to 
near--infrared (F160W) bands. The importance of such a study is
related to the recent release of new classical Cepheids observed with HST/WFC3 in 8 distant
galaxies where SNIa are hosted. These observations have posed sound constraints to the
current distance scale with uncertainties on the Hubble constant H$_0$
smaller than 3\%. Our models explore a large range of metallicity and Helium content, thus providing a robust and unique
theoretical tool for describing these new and future HST/WFC3
observations. As expected, the Period--Luminosity
(PL) relation in F160W filter is linear and slightly dependent on the metallicity when compared with optical
bands, thus it seems the most accurate tool to constrain extragalactic
distances with Cepheids.\par
We compare the pulsation properties of Cepheids 
observed with HST/WFC3--IR with our theoretical scenario and we
discuss the agreement with the predicted 
Instability Strip for all the investigated galaxy samples including
the case of NGC4258. \par
Finally, adopting our theoretical F160W PL relation for Z$=$0.02 and
log P$\gsim$1.0, we derive new distance moduli. In particular, for NGC
4258, we derive a distance modulus $\mu_0 =$ 29.345 $\pm$ 0.004 mag with a $\sigma =$ 0.34 mag, which is in
very good agreement with the geometrical maser value.
Moreover, using the obtained distance moduli, we estimate the
Hubble constant value, H$_0=$76.0$\pm$1.9 km s$^{-1}$ Mpc$^{-1}$ 
in excellent agreement with the most recent literature values.
\end{abstract}

\begin{keywords}
galaxies: spiral
---stars: distances
---stars: variables: Cepheids 
\end{keywords}

\section{Introduction}
\label{intro}
Major improvements in the classical Cepheid extragalactic distance
scale have been obtained in the 
last 20 years thanks to the Hubble Space Telescope (HST)
observations of a number of galaxies containing both Cepheids and
SNIa out to distances of $\sim$25 Mpc \citep{freedman10}.
In the literature, the Cepheid period--luminosity (PL)
relation is often adopted in the optical
bands (V and I) and calibrated using the relation of the
Large Magellanic Cloud (LMC) Cepheid sample. However, the optical PL
relations suffer from metallicity effects and, in the case of metal
poor samples (Z$<$0.01), nonlinearity \citep[see
e.g.][]{caputo00}. Moreover, the zero point depends on the assumed LMC distance and the slope is traditionally assumed to be
universal, in spite of several investigations that find a non negligible metallicity effect \citep{sakai04,tammann03,bono10b,ngeow12}. 
The LMC has the advantage of being quite near (about
50 Kpc), so that its distance has been measured using different
primary distance indicators \citep[RR Lyrae, Tip of the Red Giant Branch, Red--Clump,
etc,][]{fiorentino11,walker12,laney12,haschke12,marconi12}.
Moreover, a large sample of Cepheids has been released by the
micro--lensing experiments carried out in this galaxy. In particular, the last
catalogue from OGLE III includes more than 3,000
objects \citep[][]{soszynski08b}. 
However, the HST galaxies with SNIa typically have metallicities much
higher than the LMC (Z$>$0.02 compared to Z$=$0.008). The metallicity
dependence of the PL relation has been largely debated in the literature both from the theoretical and the observational
point of view, even if a general consensus has not been achieved yet
\citep[e.g. ][and references
therein]{kennicutt98,caputo02,sakai04,macri06,fiorentino02,marconi05,bono08,bono10b,romaniello08}. Nonlinear
convective pulsation models also suggest a non negligible dependence on
the Helium content \citep{fiorentino02,marconi05}.

\begin{figure*}
\centering
\includegraphics[width=8.5cm]{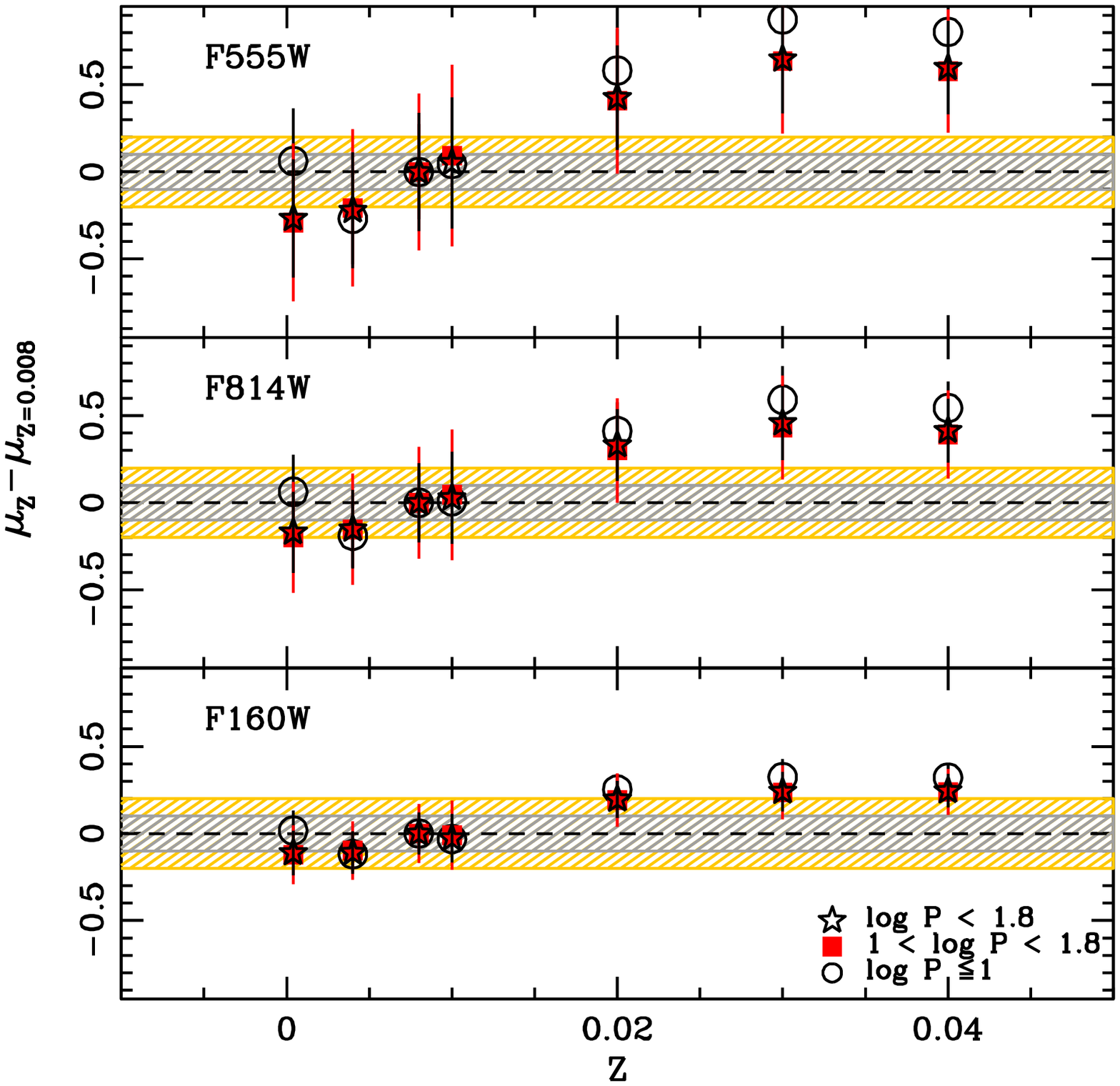} 
\includegraphics[width=8.5cm]{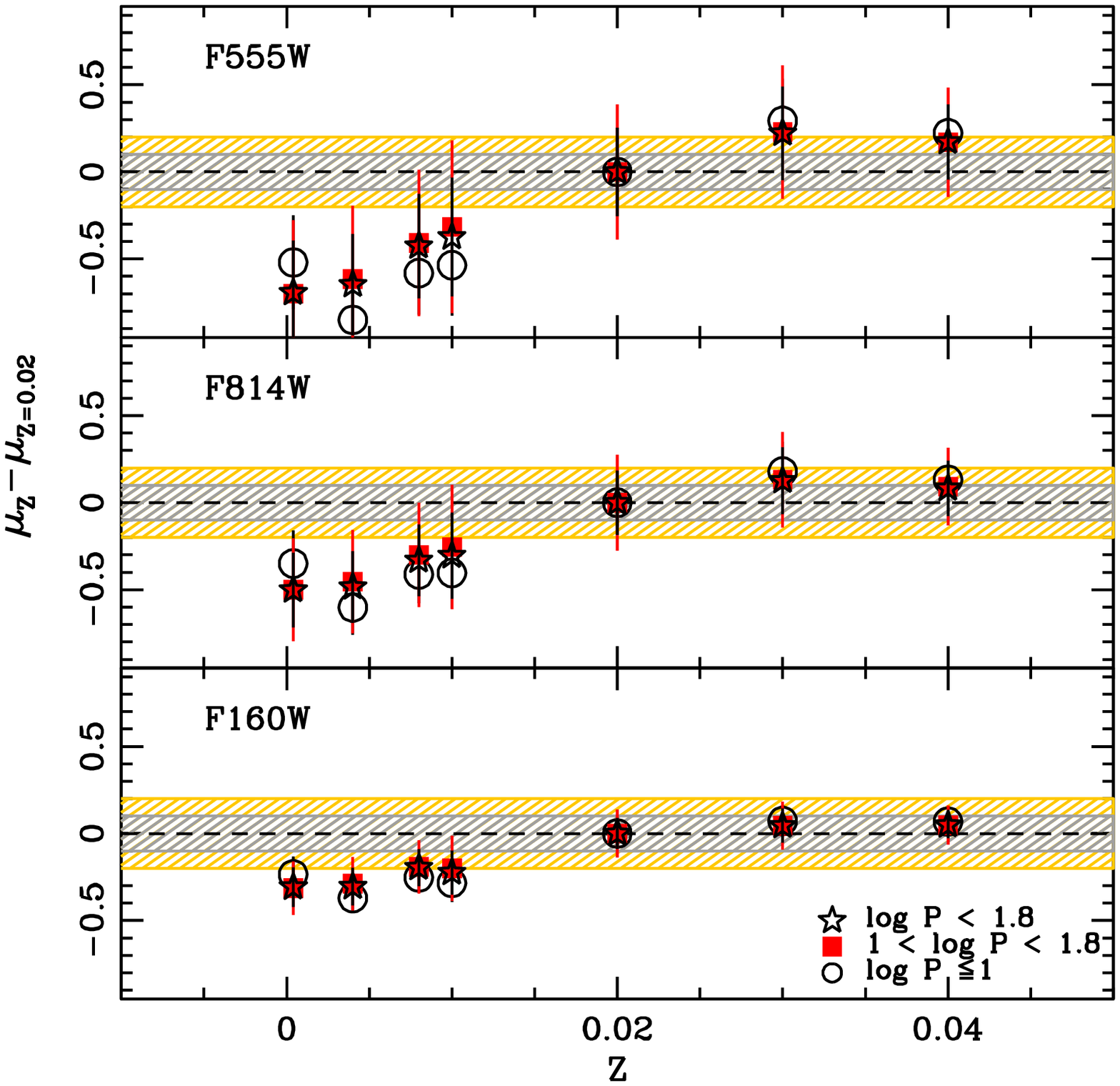} 
\caption{Difference between the theoretical distance moduli obtained
  for each selected metallicity and the corresponding value at
  Z$=$0.008 (left) and Z$=$0.02 (right). This effect is estimated using ``broken
linear'' PL relations in three filters, as labeled in the panels. Grey and light
  orange regions define 0.1 and 0.2 magnitude range and correspond to
  errors at the level of 5\% and 10\% in the final value of the Hubble
  constant H$_0$. Different period
  ranges, namely log P $\le$ 1 (open circles), log P $<$ 1.8 (open
  stars) and 1 $<$ log P $<$ 1.8 (red squares) are shown.}
\label{fig1}
\end{figure*}

Moreover, there are suggestions that the optical PL relations in the SMC and LMC are not linear over the whole period range, showing a change in the slope at periods around 10 days \citep[see][for details]{sakai04,ngeow05,marconi05}. This does not seem to be the case for Galactic Cepheids \citep{tammann03}. Although further investigations on the nonlinearity are needed for definitive claims,
LMC might not be the ideal sample to set both the slope and the
zero--points of optical PL relations.

On the other hand, observational and theoretical
investigations suggest that in the 
near--infrared (NIR) bands (J and K), that are also much less affected
by the reddening,
both the chemical composition and nonlinearity effects are
significantly reduced
\citep{bono99a,marconi05,fiorentino07,freedman12}. Another advantage of
the NIR bands is that, given the small pulsation amplitudes in these
filters, only one epoch is sufficient to
derive accurate mean magnitudes (with an error of about 0.05 mag), once
detections and periods are available from optical data \citep[][]{soszynski05,scowcroft11,ripepi12,inno13}.
In this context, \citet{riess11a} decided to follow-up with the WFC3-IR camera
on board HST the Cepheid samples discovered in seven SNIa hosting
galaxies previously observed in the optical bands by using
WFPC2 and/or ACS@HST \citep{riess09a,riess09b}. 
To further minimize the uncertainties in the zero point due to the
metallicity difference among LMC and metal--rich galaxies,
\citet{riess11a}  have chosen NGC4258 as anchor for the Cepheid PL
relation. In fact, this galaxy has a very accurate geometrical
distance estimate, d$=$7.2$\pm$0.5 Mpc \citep[or $\mu_0
=$29.29$^{+0.14}_{-0.16}$ mag][]{herrnstein05}, based on
the analysis of its H$_2$O maser emission.
To remove the reddening uncertainty, these authors adopt a Wesenheit
relation based on the combination of the three filters $V$, $I$, and
$F160W$ \citep[see][for details]{riess11a}, in the
assumption that the extinction law is universal.

To properly compare our set of nonlinear convective pulsation models
\citep[see][and references therein]{marconi05,bono08} to these
recent observations we have transformed the whole framework in
WFC3/HST filters. 
The paper is organised as
follows. Section 2 briefly describes the theoretical framework and
its physical ingredients. The multi--wavelength relations connecting
periods with magnitudes and colours are presented in
Section 3. In Sections 4 and 5 we discuss a
comparison between theory and the Cepheid samples observed in
SNIa hosting galaxies. In Section 6 we give new estimates of the
distance moduli for these galaxies and, correspondingly, a new value
for the Hubble constant H$_0$. The conclusions close the paper.  

\section{Pulsation models and transformations}

In the last 15 years, we have computed an extended set of pulsating models for masses
representative of Classical Cepheids (3$\lsim$M/M$_{\odot}$$\lsim$12)
using a code originally developed by \citet{Stellingwerf78} and
subsequently adapted to Classical Cepheids by \citet{bono99b,bono99a}.
A large range of input physical parameters has been explored in order
to investigate their effect on the Cepheid intrinsic properties \citep{bono00a,fiorentino02,marconi05,fiorentino07,marconi10}.
In particular, the adopted metallicities range from Z$=$0.0004 to Z$=$0.04, while the Helium content
(depending on the assumption on the $\Delta Y$/$\Delta Z$ ratio)
ranges from 0.24 to 0.34. 

With this set of nonlinear pulsational models we predict both the red
and the blue boundaries of the classical instability strip (IS) in the
Hertzprung--Russel diagram (log L/L$_{\odot}$ vs log T$_{eff}$), for
each assumed chemical composition. We also build a synthetic stellar
population filling the predicted IS with masses in the range 3 -- 12 M$_{\odot}$ and with an initial mass function that follows the relation
m$^{-3}$. Adopting theoretical
evolutionary prescriptions, we can associate to each synthetic stellar
mass a luminosity given by a canonical\footnote{The canonical
  Mass--Luminosity relation is based on evolutionary models without
  core over-shooting and mass--loss, as discussed in \citet{bono00a}.} Mass--Luminosity
relation whereas the effective temperature is randomly extracted
within the IS for a given luminosity level.\par
These synthetic Cepheid distributions can be transformed into the
WFC3/HST photometric system. To this aim, we use the bolometric corrections for
both UVIS and IR channels at WFC3/HST carefully computed by S. Cassisi and
A. Pietrinferni (priv. comm.). In the following we focus on the
most commonly used filters, i.e. F555W (narrow V), F606W (wide V), F814W (wide I) and F160W
(wide H). The transformed scenario in the remaining filters is available upon requests. To
transform theoretical luminosity and effective temperature into
magnitudes and colours we have assumed an updated value for the solar bolometric
magnitude M$_{bol,\odot} = $ 4.77 mag (S. Cassisi, priv. comm.). 
Through regression of these synthetic populations we can finally derive multi--wavelength relations, such
as the PL, the PL--Colour (PLC) and the Wesenheit relations.

\section {Theoretical PL and Wesenheit relations}\label{rel_teo}
In this section we analyse the properties of the theoretical PL, PLC and Wesenheit relations
 in the WFC3/HST photometric system with particular attention to the NIR filters, given the
increasing role played by both space and ground--based NIR
facilities.

We derive the linear PL relations for the whole set of available
chemical compositions whose coefficients are reported in
Table~\ref{tab1}. As expected, these relations become steeper and tighter when passing
from optical (F555W) to NIR (F160W) bands with the scatter being reduced by
a factor 2. These results confirm that NIR PL relations provide more
accurate results than the optical ones , as found recently for the LMC Cepheids
\citep{soszynski08b,ripepi12,inno13}. This finding is further
supported by the significant reduction of the reddening effect in
these bands, as the ratio between the total to selective
extinction decreases of about a factor 5 going from optical to NIR
filters \citep[R$_{F555W} \sim$ 3.06 and R$_{F160W} \sim$
0.41, derived from][]{cardelli89}.\par

Given the quoted non linearity of the optical PL relations, at least
for metallicity typical of the Magellanic Clouds, we derive the
coefficients of both the quadratic relations and the linear ones
(``broken linear'' PL) for
periods shorter and longer than 10 days (see Table~\ref{tab2}). 
As expected, with the exception of the very
low--metallicity case (Z=0.0004), the coefficient of the quadratic term increases by
decreasing the metallicity, but the effect is almost removed in the
NIR bands. 

Since a general consensus on the chemical composition effect on PL
relations has not been achieved yet, the metallicity dependence
contribution to the final error budget on the extragalactic distance
scale is not uniquely quantified. In this context, on the basis of
the relations derived in this paper, we provide new constraints to the
metallicity effect. 

\begin{figure}
\centering
\includegraphics[width=8.5cm]{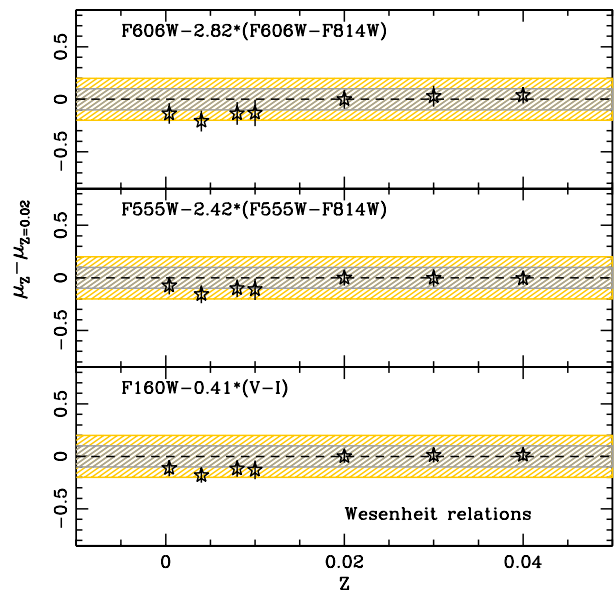} 
\caption{As in Fig.~\ref{fig1}, but using the Wesenheit relations
    for the three colour combinations labeled in literature and using
    as reference the solar--like metallicity $Z=0.02$.}
\label{fig2}
\end{figure}

In Fig.~\ref{fig1}, by using our synthetic Cepheid samples, we show the predicted differential distance
moduli relative to the reference model set with metallicity Z$=$0.008
(LMC) and Z$=$0.02 (sun) as a function of the metal content. The
distance moduli are obtained with ``broken linear'' PL relations in the labelled filters. We have distinguished three different
period ranges.

As expected, the effect is more significant in the optical
filters. Moreover, inspection of this figure suggests that the
adoption of the correct reference metallicity is crucial for a significant
reduction of the final error in the extragalactic distance scale.  In
particular, for the supersolar metallicities of the HST galaxies, the
adoption of PL relations for solar metallicity reduces the error
due to the metallicity effect to few percent.  This evidence supports
the use of PL relations calibrated on NGC 4258 as adopted by \citet{riess11a}.
Considering this result and the period range typical of these
extragalactic Cepheid samples (see below),
in the following, we adopt the theoretical PL
relations for Z$=$0.02 and Y$=$0.28  derived for log P $>$ 1.

\begin{figure*}
\centering
\includegraphics[width=8.5cm]{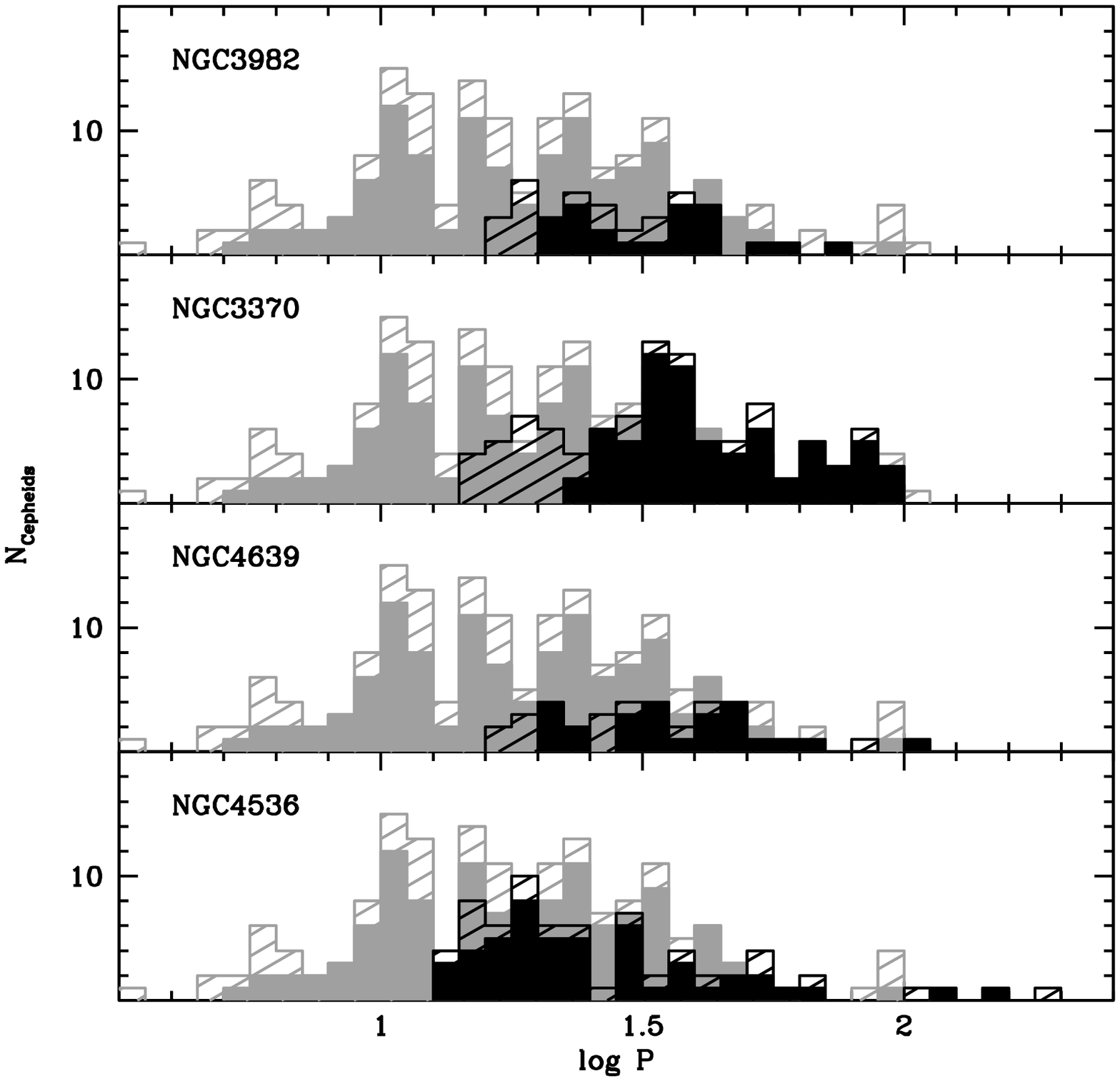} 
\includegraphics[width=8.5cm]{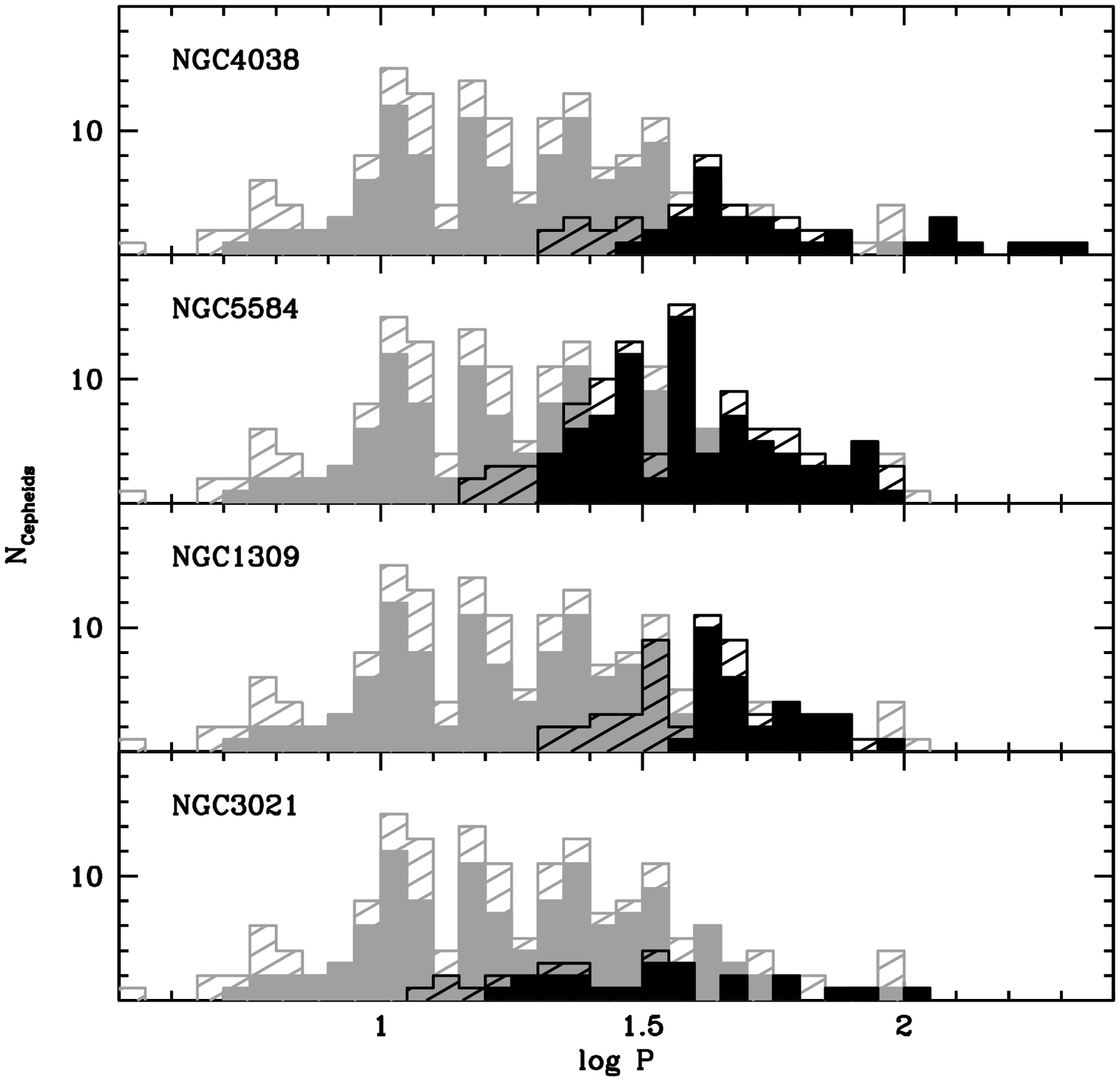} 
\caption{Period distributions of the eight host galaxy (black) compared to
  NGC4258 (grey). The name of the galaxy is labeled in each panel. The
  shadow regions are the whole Cepheid sample, the filled regions
  represent the {\it good} Cepheids used by \citet{riess11a}.}
\label{fig3}
\end{figure*}

One of the most commonly used tool to derive Cepheid distances  is the Wesenheit
relation \citep{madore76,fiorentino07,bono08,fiorentino10a}. This
formulation has the advantage to be reddening$-$free by
construction and to include a colour term, and in turn the temperature
information, for each individual star, defining a
very tight relation. In Table~\ref{tab3} we show some of the classical
formulations together with the new recent
one that combines Johnson-Cousin optical bands and the NIR ones from WFC3/HST
photometric system \citep{riess09a,riess11a}. These relations have
been derived using the following extinction coefficients:
R$_{F160W}$=0.41, R$_{F814W}$=1.79, R$_{F606W}$=2.78 and R$_{F555W}$=3.06 from \citet{cardelli89}.
The predicted intrinsic dispersions are very small, as well as the
metallicity effects, making these
relations powerful tools for accurate distance determinations.

In Fig. ~\ref{fig2} we show the predicted differential distance
moduli relative to the reference model set with metallicity Z$=$0.02
(sun) as a function of the metal content obtained using the Wesenheit
relations. This Figure show that the metallicity effect with such
relationships is very small, i.e. $\le$0.2 mag or $\le$10\% in the final value of the Hubble
  constant H$_0$.

\section{Comparison with the WFC3@HST Cepheid sample}

As we discussed in the introduction, one of the aims of this paper is the
comparison of our pulsation framework with the large
Cepheid sample recently observed in NGC4258 and in 8 far SNIa host
galaxies with WFC3@HST \citep{riess11a}. In order to
provide new, solid, constraints to the Hubble constant H$_0$,
\citet{riess11a} classified Cepheids in each galaxy as {\it good} or
{\it rejected} through a sigma--clipping (at 2.5 $\sigma$) rejection criterium with
respect to the F160W--PL relation defined for NGC4258
Cepheids \citep[see][for details]{riess11a}. \par
Table~\ref{tab4}  shows the global properties of these galaxies,
namely the galaxy ID, the number of total vs {\it good} Cepheids
observed in each galaxy, the mean Cepheid metallicity, the
SNIa apparent magnitude and their distance modulus. For each
Cepheid, coordinates, period, F160W magnitude and V-I
colour are available in \citet{riess11a}. Only for few galaxies (NGC4258, NGC3370,
NGC3021 and NGC1309) optical (V, I) individual magnitudes are also
at disposal in the literature \citep{macri06,riess09a,riess09b}.\par

\subsection{The Cepheid period distributions}

\begin{figure*}
\includegraphics[width=8.5cm]{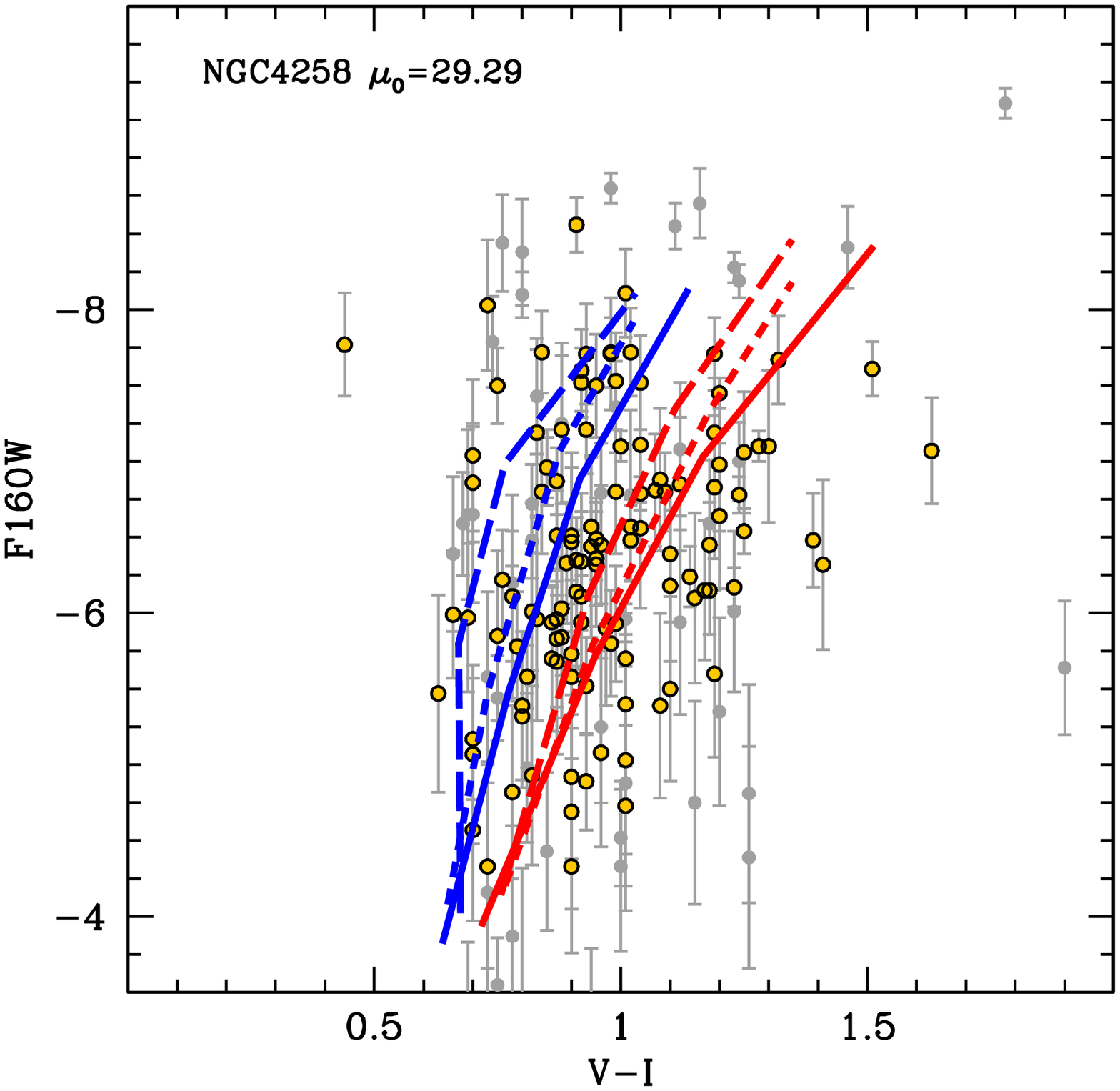} 
\includegraphics[width=8.5cm]{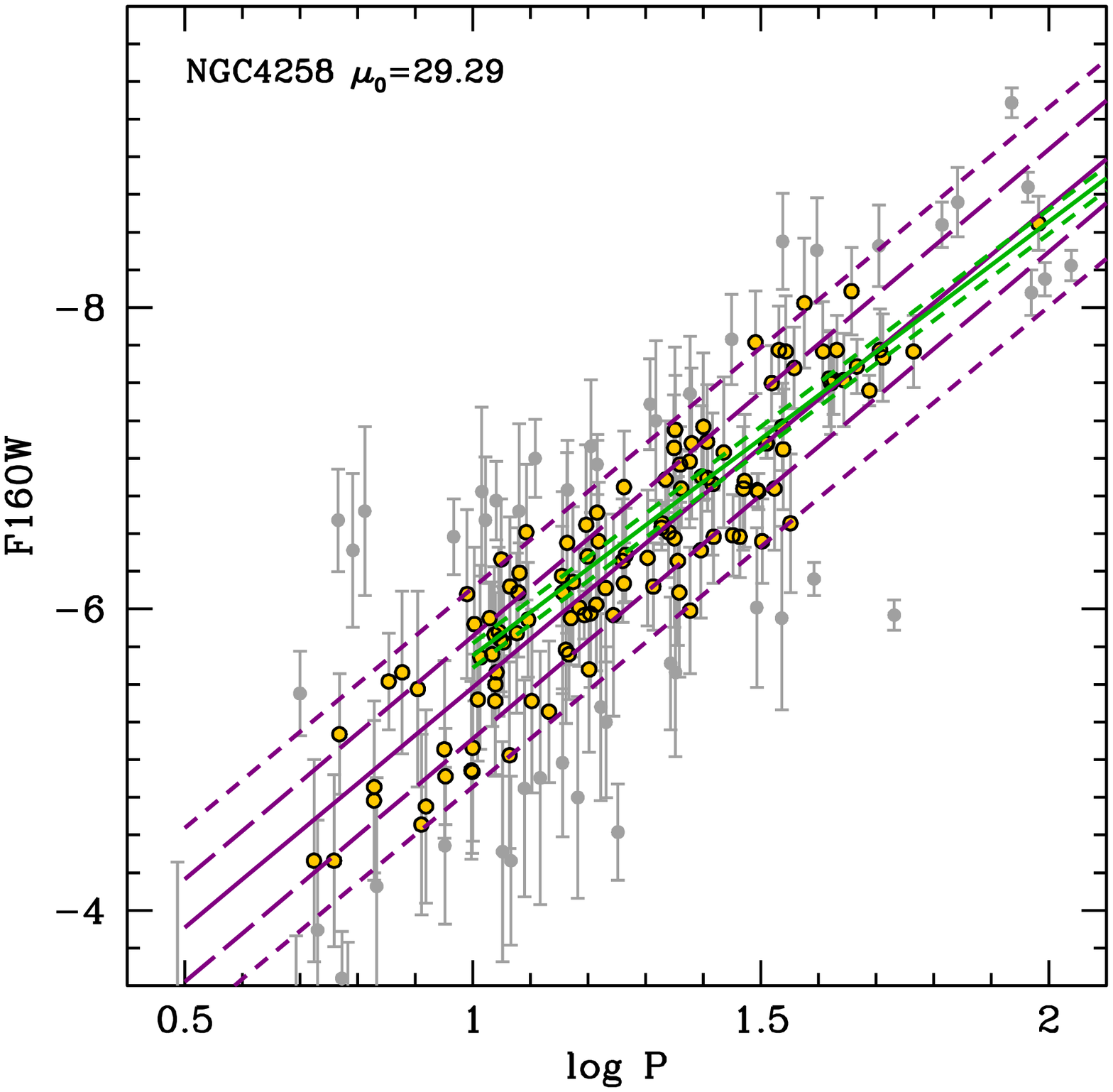} 
\caption{--{\it Left}-- Comparison between the observed magnitude--colour
  distribution (F160W, V--I) of NGC4258 Cepheids and the
  theoretical boundaries of the pulsation Instability Strip for
  three different assumptions on the chemical composition:
  Z=0.02;Y=0.28 (solid), Z=0.02;Y=0.31 (dashed) and Z=0.008;Y=0.25
  (long dashed). Orange and grey Grey symbols have been used for good
  and rejected observed Cepheids, respectively. --{\it Right}-- The
  same as in the left panel but
  in the F160 vs logP plane. In this panel, we also show the
  theoretical (Z$=$0.02, Y$=$0.28 and log P $>$ 1; dark green solid line) and the observed (purple solid line)
  F160W--PL relation together with their confidence line (dashed for
  the whole Cepheid sample and long-dashed for the good ones).}
\label{fig4}
\end{figure*}

In Fig.~\ref{fig3}, we plot the comparison among the period distributions of the various
galaxies, including the anchor galaxy NGC4258. As well known, the Cepheid samples observed by HST in very distant galaxies
are affected by an observational bias both at the short and long-period end of the
PL relation due to the limited observational
time baseline \citep[see e.g.][and references therein]{Freedman01}. 

Inspection of Fig.~\ref{fig3} confirms that only for the anchor galaxy
NGC4258 periods cover the short range \citep{macri06}, whereas for all the
others the period distribution span a quite similar range
($\sim$12--100 d), with a few significantly ultra long period Cepheids in
NGC4536 and NGC4038 \citep[][]{fiorentino12b}. 

\subsection{The Cepheid Instability Strip and F160W--PL relation}\label{strip}

In this section, we compare our pulsation theoretical predictions to
the observations. In Fig.~\ref{fig4} (left panel) we show the
theoretical IS compared with the Cepheid distribution in NGC4258 in the magnitude (F160W) vs colour (V--I) plane. The theoretical ISs for fundamental pulsators are
shown for three selected chemical compositions: the solar--like Z$=$0.02;Y$=$0.28, the
Helium enriched Z$=$0.02;Y$=$0.31 and the LMC--like
Z$=$0.008;Y$=$0.25. We assume the distance modulus $\mu_0 =$29.29 mag by
\citet{herrnstein05} and the extinction values by \citet[][see columns 7-8-9 in  Table~\ref{tab4}]{schlafly11}. We use different symbols to
indicate {\it good} (light orange dot) and rejected
(grey dot) Cepheids. We note that the observed Cepheids 
show a significantly larger
spread than predicted by pulsational
models for the solar composition (solid lines). Increasing the Helium
abundance, the IS boundaries (dashed lines) move towards hotter
effective temperatures (bluer colours). A still larger effect is found
for metal--poor LMC--like Cepheid IS boundaries
(long-dashed lines). In any case, on average, the colour distribution of Cepheids
in NGC4258 seems to be better reproduced by the two model sets with
Z$=$0.02.  \par

This occurrence is still evident when moving to the NIR PL (right
panel of Fig.~\ref{fig4}), where pulsation models predict a very
narrow and linear relation (dark green lines). This prediction is well supported by recent NIR observations of
Cepheids in the LMC and in our Galaxy
\citep{ripepi12,inno13,freedman12}. From these observations, it is evident
that, for  wavelength going from
$\sim$1.2 (J) to 3.6 (mid IR) micron \citep{scowcroft11,monson12}, the measured scatter of the PL relation is always less than
$\sim$0.12 mag. Instead, a linear regression of all the data in right panel of
Fig.\ref{fig4} (purple lines) provides a $\sigma_{PL} =$0.65 mag. From a further inspection of this Figure, we also note that theoretical models predict a flatter slope
($\beta_{(Z=0.02;Y=0.28)} = -$2.88) than observed ($\beta_{(NGC4258)} = -$3.24) and a larger zero point
($\alpha_{(Z=0.02;Y=0.28)} = -$2.81 vs $\alpha_{(NGC4258)} = -$2.20).\par

\begin{figure*}
\centering
\includegraphics[width=16cm]{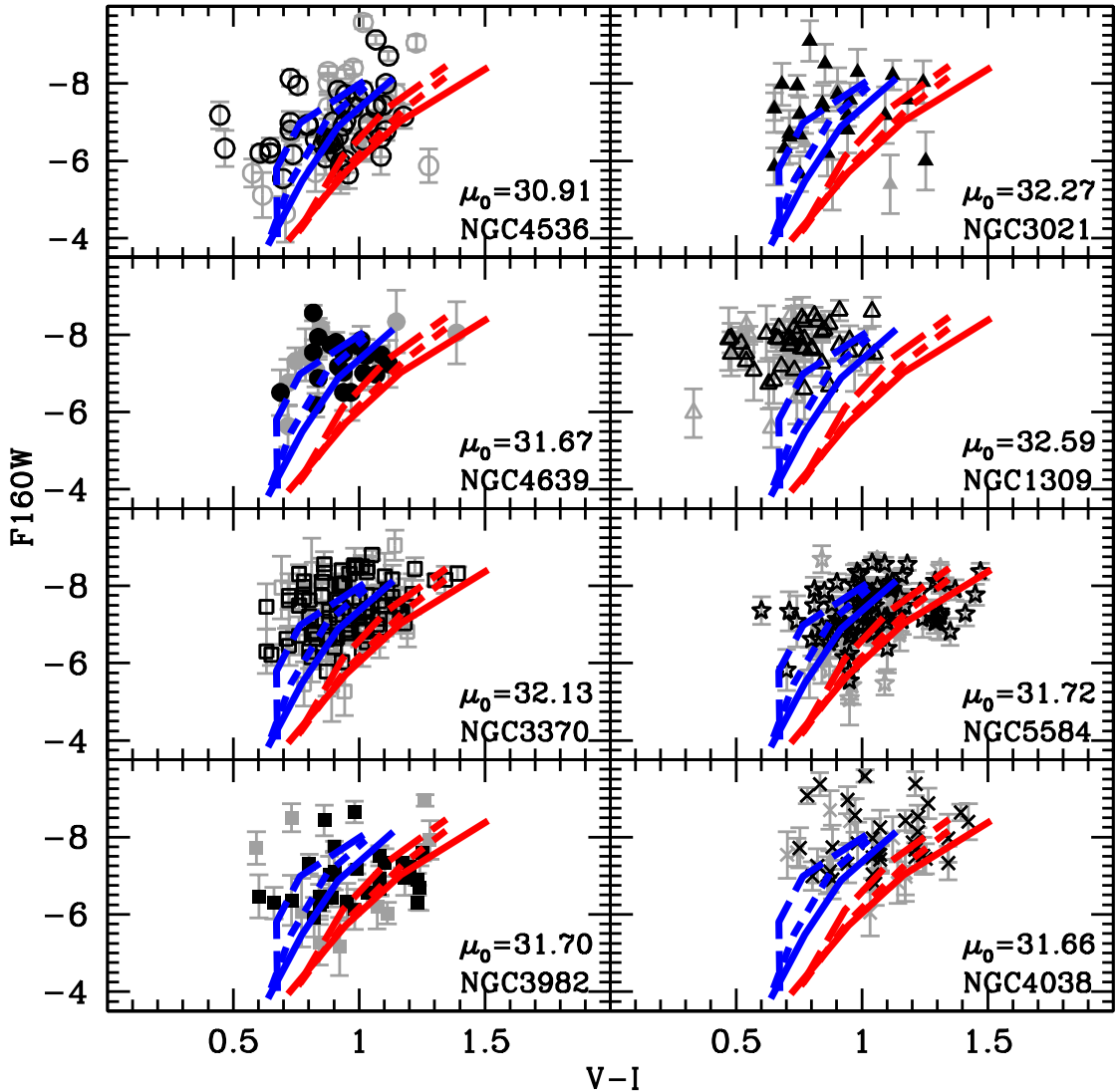} 
\caption{The same comparison as in Fig.~\ref{fig4} (left), for each galaxy of the
  sample. In this case, black and grey symbols represent good and
  rejected Cepheids as selected in \citet{riess11a}, respectivley.}
\label{fig5a}
\end{figure*}

\begin{figure*}
\centering
\includegraphics[width=16cm]{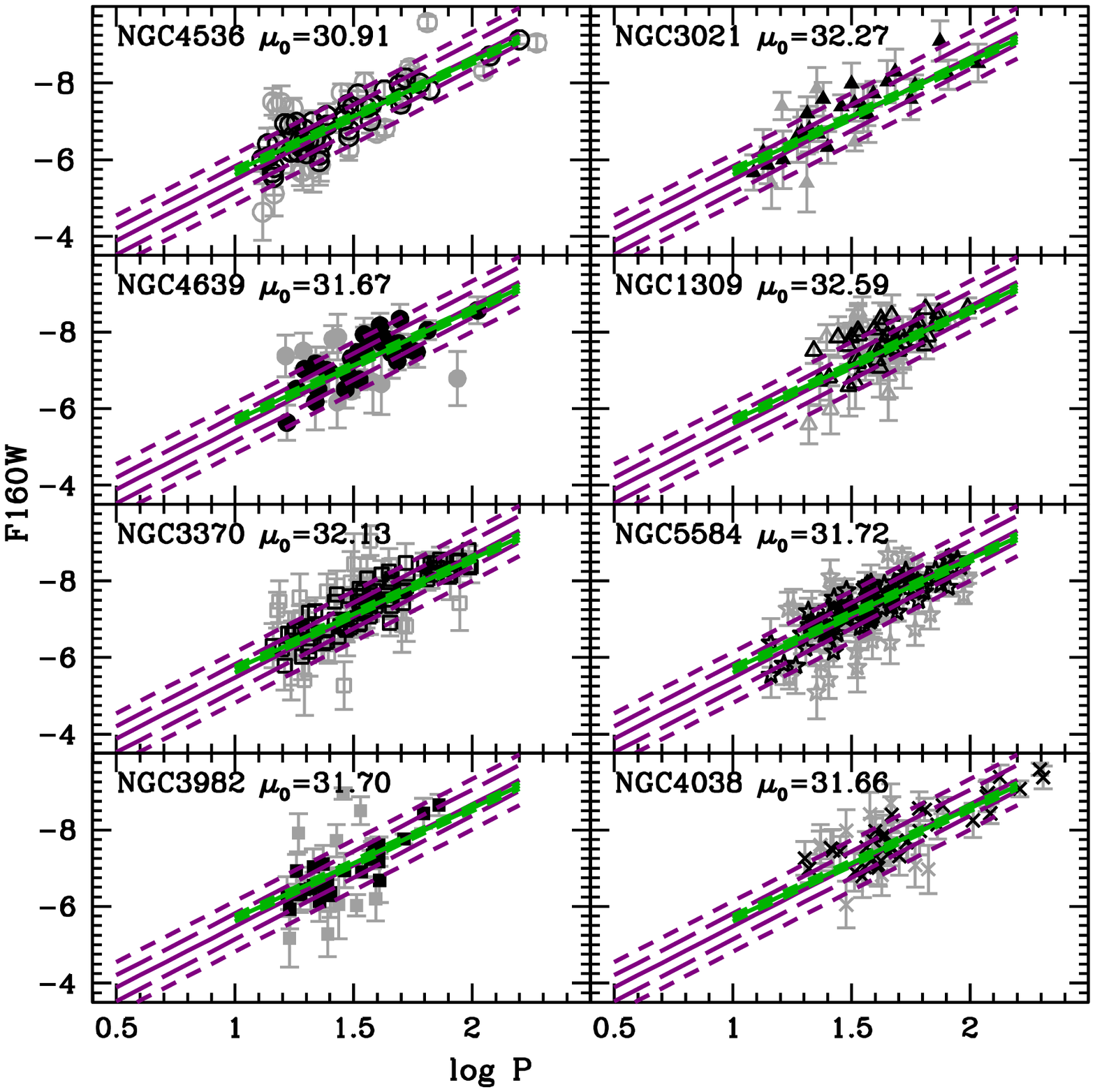} 
\caption{The same comparison as in Fig.~\ref{fig4} (right), for each galaxy of the
  sample. In this case, black and grey symbols represent good and
  rejected Cepheids as selected in \citet{riess11a}, respectivley.}
\label{fig5b}
\end{figure*}

In Fig.s~\ref{fig5a}-\ref{fig5b} (left panel) we show the CMD location of Cepheids for each
HST sample. Also for these galaxies, we notice a large spread
in the observed distributions when
compared with the predicted IS boundaries. Moreover, the Cepheid location is on average bluer and
brighter than theoretical predictions, with the effect increasing for
more distant galaxies. These blue colours have been already
pointed out by \citet{tammann12} for the {\it extreme} cases of
NGC1309 and NGC3021 when compared with Galactic Cepheids. 
Our pulsational framework suggests very blue colours only for very
metal--poor and  Helium enriched environments. However, a large
variation in the chemical abundance is not supported by recent estimates of the mean metallicity [O/H] in
each galaxy (see column three of Table~\ref{tab4}). We notice that
an overestimate of the reddening can not be the culprit for these blue
colours, because the same trend is observed even if we do not account at
all for the Galactic extinction. In the right panel of the  Fig.s~\ref{fig5a}-\ref{fig5b}, we show the F160W PL 
distribution for each galaxy. 
As in the case of NGC4258, in spite of the large spread of
these distributions, the agreement with theoretical F160W PL relations
is quite satisfactory in particular for {\it good} Cepheids (black symbols).\par

The puzzling colour/magnitude behaviour shown by Cepheids could depend on the well
known difficulties to perform accurate stellar photometry in 
crowded distant galaxies. This problem worsens as the distance
increases. In particular, a possible systematic effect can be related
to the blending from unresolved nearby companions \citep[see e.g.][and
references therein]{stetson98,chavez12} that might produce the observed shift to the
hotter and slightly brighter region of the CMD. 

\section{Comparison with the ACS@HST Cepheid sample} 

In this Section we focus on the comparison between our theoretical
scenario and the properties of Cepheids detected in
NGC4258, NGC1309, NGC3021 and NGC3370. 
For these galaxies individual Johnson V, I magnitudes have been also derived from observations in the  F555W and F814W filters of
the WFC@ACS on board HST \citep{macri06,riess09a,riess09b}.

\begin{figure*}
\centering
\includegraphics[width=8.5cm]{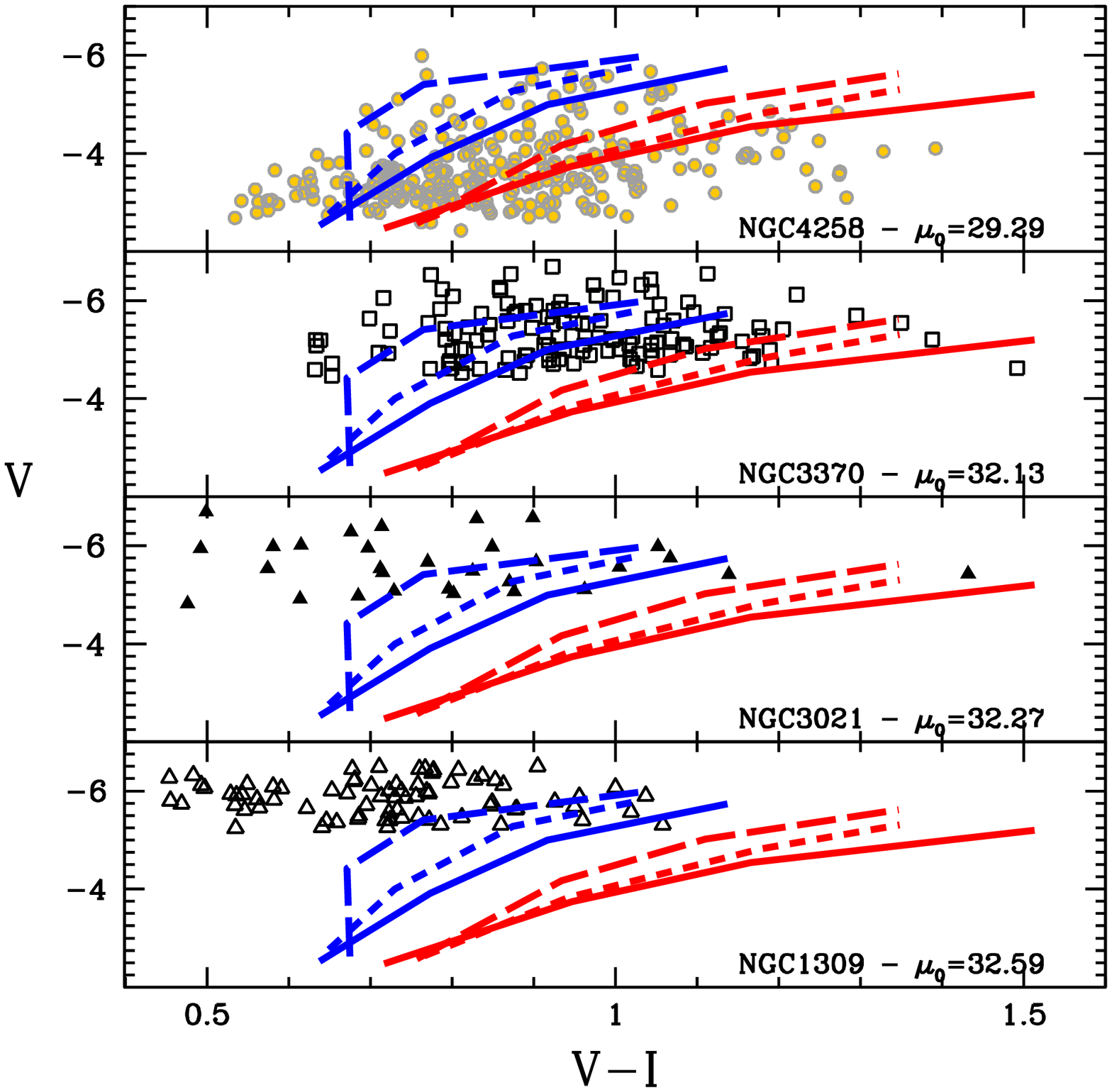} 
\includegraphics[width=8.5cm]{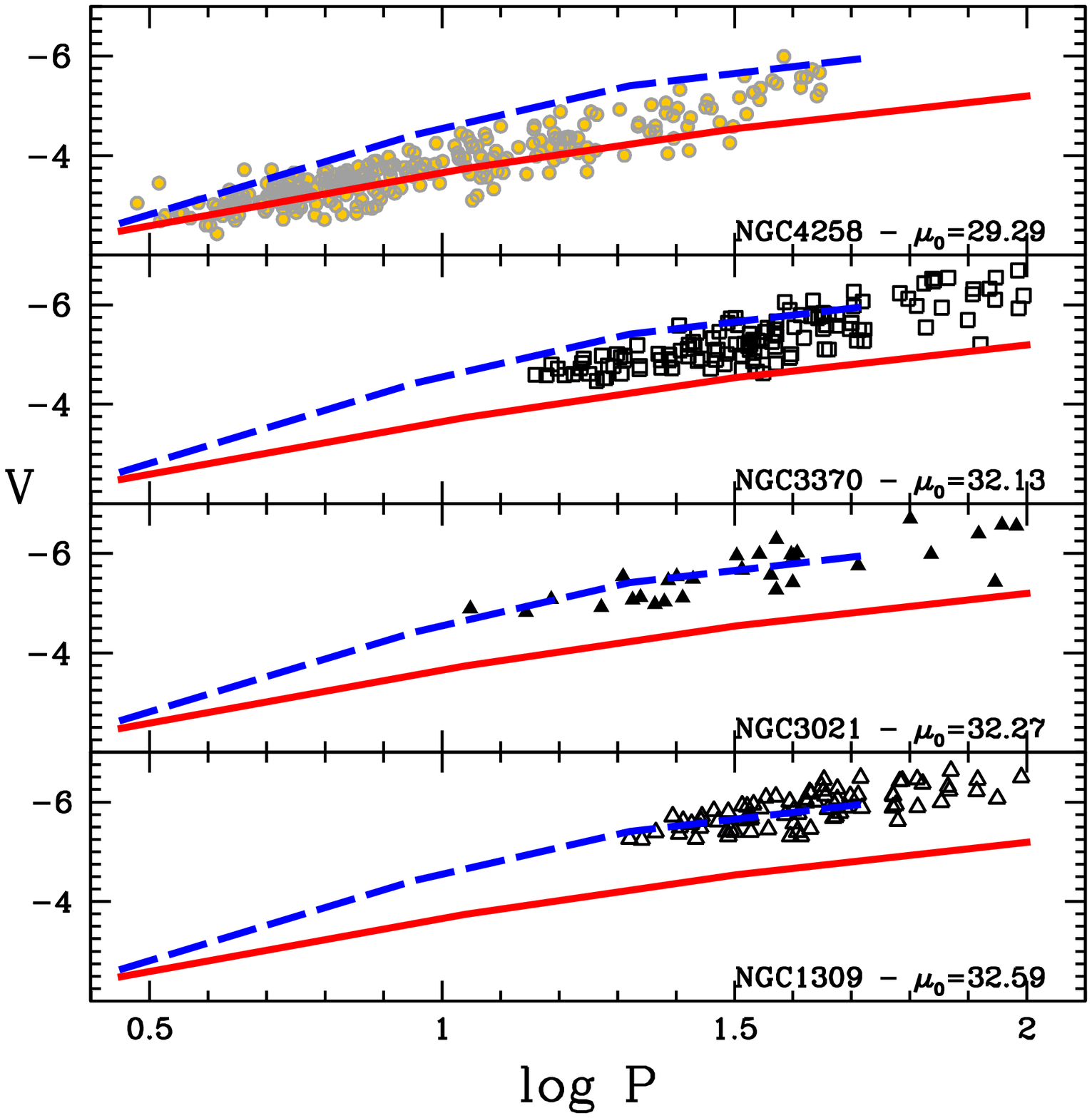} 
\caption{ --{\it Left} -- Comparison between the observed magnitude--colour
  distribution (V, V--I). The theoretical boundaries of the
  pulsation Instability Strip are plotted for the same chemical compositions
  chosen in Fig.s~\ref{fig5a}-\ref{fig5b} as well as the same symbols for each
  galaxy are used.--{\it Right} -- The V Period--Luminosity distribution compared with theoretical
prediction. For comparison, we have over--plotted the blue boundary of the IS for Z = 0.008 (LMC--like) and the red boundary for Z = 0.02 (solar--like).}
\label{fig6}
\end{figure*}

In Fig.~\ref{fig6} (left panel) we show a comparison between
the observations and the theoretical
boundaries of the ISs for fundamental pulsators, for the three
chemical compositions adopted in Fig.s~\ref{fig5a}-\ref{fig5b}.
We have ordered the galaxies from top to bottom by increasing the
distance modulus. As already observed in the F160W vs
V$-$I plane, as the galaxy distance increase, the Cepheids
location in the CMD moves to bluer colours and brighter magnitudes.
A similar trend for the Cepheid brightness is noted in the PL plane
(right panel of Fig.\ref{fig6}).
For comparison, we
have over-plotted  the blue boundary of the fundamental IS for
$Z=0.008$ (LMC-like) and the red boundary for $Z=0.02$
(solar-like) to take into account a possible metallicity spread  found
in some of these galaxies \citep{bresolin11}. In this plane, the
agreement between theory and observations improves thanks to a smaller
dependence on reddening and uncertainties in the colour--effective temperature transformations.
In the case of  NGC4258, when its metallicity spread is
accounted for, the comparison with the predicted IS is
satisfactory. The same holds for NGC3370, whereas the other two
galaxies show a Cepheid distribution that is significantly bluer than
predicted, even taking into account a possible revision of their metal
abundance toward $ Z\sim0.01$ as suggested by \citet{bresolin11}. 

There is currently no clear explanation in the literature for the
blue shift observed in distant galaxies, as well as for the very
large color spread displayed by all the investigated samples
\citep[see also][for a discussion]{tammann12}. A possible
contribution to this trend could be given by a blending effect  with faint
unresolved companion stars that is expected to increase with the galaxy distance.

\section{New distances and their impact on H$_0$}\label{new}

In this section, we derive new distance moduli for the galaxies of the sample using our set of
pulsation models.
Although the use of the Wesenheit relation is very common to derive
the distance moduli and usually
provides smaller errors than multi--wavelength PL both from
observational and theoretical point of view, due to the discussed
$V-I$ uncertainties in these HST data, we prefer to use a pure NIR
F160W--PL relation . \par

In particular, we adopt the following theoretical F160W
PL for Z=0.02 (Y=0.28) and log P $>$ 1.0, given the period range
available for distant galaxies.

\begin{equation}
F160W = - (2.81 \pm 0.08)  - (2.88 \pm 0.01) log P  
\end{equation}

The corresponding coefficients are also highlighted in bold--face in
Table~\ref{tab2} and taking into account the small slope difference between the theoretical F160W
PL (for Z=0.02 and Y=0.28 and log P $>$ 1) and the one by
\citet{riess11a}, we adopt their selection of {\it good} Cepheids.

As a first check, we applied our theoretical relation to the {\it good}
Cepheids with log P $>$ 1 in NGC4258 finding
$\mu_0 =$ 29.345 $\pm$ 0.004 mag with a $\sigma =$ 0.34 mag, which is in
very good agreement with the maser distance: $\mu_0=$29.29$^{+0.14}_{-0.16}$ mag.

\begin{figure}
\centering
\includegraphics[width=8.5cm]{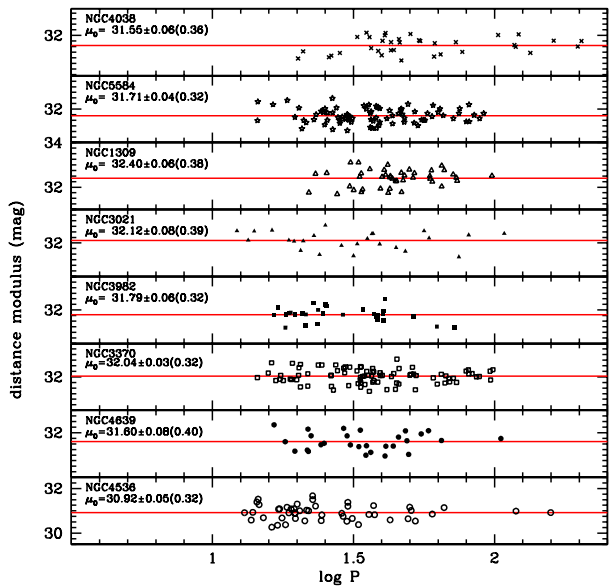} 
\caption{Individual Cepheid distance moduli obtained with the
  theoretical F160W PL relation
  for Z$=$0.02,
  Y$=$0.28 and log  P $>$ 1.  In each panel, the solid line represents the inferred mean
  value. }
\label{fig7}
\end{figure}

\begin{figure}
\centering
\includegraphics[width=8.5cm]{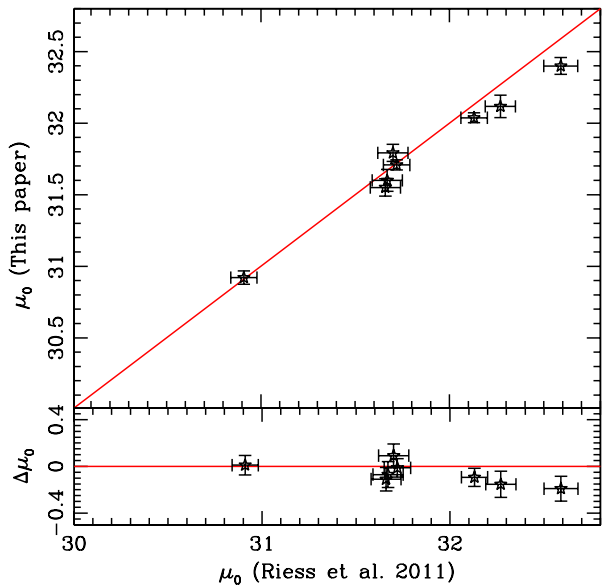} 
\caption{--{\it Top panel}--Comparison between the distance modulus found by
  \citet{riess11a} using a metal dependent Wesenheit relation
  and the one derived in this paper using our theoretical
  F160W PL for Z$=$0.02--Y$=$0.28 and for periods
  larger than 10 days. --{\it Bottom panel}-- Difference of the
  distance moduli as a function the distance modulus found by
  \citet{riess11a}.}
\label{fig8}
\end{figure}

Then, we applied the same relation to the {\it good} Cepheids of the other galaxies, obtaining
the individual Cepheid
distance moduli and the resulting mean values, as plotted in Fig.~\ref{fig7} and listed in column 7 of
Table~\ref{tab4}. These values are derived adopting a reddening
correction \citep[][]{schlafly11} but without
assuming any metallicity dependence. As shown in
Fig.~\ref{fig7}, the Cepheid distribution around the derived mean 
distance modulus for each galaxy has a quite large scatter ($\sigma \gsim$ 0.5
mag). However, if we follow the same approach adopted by \citet{riess11a},  considering the large number of Cepheids observed in
each galaxy, we derive a much smaller final uncertainty (standard error of the mean) for each mean distance modulus.
In
Fig.~\ref{fig8}, we show the comparison between our derived mean
distance moduli and  the ones  by \citet{riess11a}. The error-bars are
relatively small because they
represent the standard errors of the means and according to them we note  a small
discrepancy at the longest distance moduli. However, the agreement is well
within 1--$\sigma$ over the whole distance range (see also Table \ref{tab4}).

On the basis of these distance moduli, we can compute the Hubble constant using the following equation:
\begin{equation}
logH_{0,i} = \frac{(m_{v,i}^0 +5 a_v) - \mu_{0,i} +25}{5}
\end{equation}
where the quantities $m_{v,i}^0 +5 a_v$ in  Table~\ref{tab4}, column 4,
have been taken by \citet{riess11a}.

We find  H$_0 =$76.0 $\pm$ 1.9 km
s$^{-1}$ Mpc$^{-1}$ ($\sigma =$ 4.8) in very good agreement within
the errors with 74.8 $\pm$ 3.0 km
s$^{-1}$ Mpc$^{-1}$  found by \citet{riess11a} on the basis of a metal
dependent Wesenheit relation, but also with their best estimation 73.8 $\pm$ 2.3 km 
s$^{-1}$ Mpc$^{-1}$ obtained taking into account different distance
scale calibrations and different error sources.
Finally, our result is also in good agreement with a completely
  independent estimation of the Hubble constant recently released by
  \citet{freedman12} that uses the Spitzer calibration to the Key Project
  sample of Cepheids, i.e. H$_0$ = 74.3 $\pm$ 2.1 (systematic)  km s$^{-1}$ Mpc$^{-1}$.

\section{Conclusions}
In this paper we present  the predictions of our nonlinear, nonlocal and time--dependent
pulsation scenario for Classical Cepheids transformed into the
WFC3@HST filters, as a unique theoretical tool to study the properties of new and future
observations with this powerful camera.
We have analysed the metallicity effect on the Cepheid PL relation in these photometric bands
and  confirmed the result already obtained in the Johnson--Cousin photometric system,
that the metallicity effect decreases when increasing the wavelength
(from F555W to F160W). In particular, the use of a NIR F160W PL relation provides
an accuracy of few percent when the correct metallicity reference
sample is adopted. \par

Our model predictions are useful to investigate the properties of
Cepheids observed in these filters in 8 distant galaxies hosting SNIa that have been recently used to estimate the Hubble
constant H$_0$. We find a satisfactory agreement between observed and
predicted Cepheid properties, 
with the exception of pulsators in NGC1309, NGC3021 and NGC3370 that
appear significantly brighter and bluer than Cepheid models. This
effect has already been
previously noted by \citet{tammann12} on the basis of
a comparison of these Cepheid samples with the Galactic one and could
be related to the increasing crowding effect at larger distances \citep{stetson98,chavez12}.\par 

We have applied the theoretical F160W PL relation for the chemical
composition currently adopted for NGC 4258 (Z$=$0.02; Y$=$0.28) and
log P $>$ 1 to
derive new individual and average distance moduli for the investigated
Cepheid samples. The resulting Hubble constant is H$_0 =$76.0 $\pm$ 1.9 km
s$^{-1}$ Mpc$^{-1}$ ($\sigma =$ 4.8), in excellent agreement with the
values obtained by \citet{riess11a} and \citet{freedman12}. This
agreement suggests that adopting a very simple and direct approach,
based on the predictions of non-linear pulsation models, we are able
to constrain the extragalactic distance scale and the Hubble constant
at the same level of accuracy obtained from an
observational point of view. \par

\bigskip
\section*{acknowledgments}

\small
The authors warmly thanks Santi Cassisi and Adriano Pietrinferni for kindly providing the updated
bolometric corrections in WFC3/HST bands for both UVIS and IR
channels. GF has been supported by the INAF fellowship 2009 grant and
by the COSMIC--LAB ERC grant (under contract ERC-2010-AdG-267675). We thank the referee V. Scowcroft for
her suggestions/comments that improved the readibility and the impact of this paper. 
\normalsize

\newpage

\begin{table*}
\centering
\caption{Linear PL relations in HST/WFC3 photometric system for a selected
  sample of filters. Other filters are available upon request.}
\begin{tabular}{llcc}
\hline
\hline
\noalign{\smallskip}
{$\rm Z$}&{$\rm Y$}&{$\rm \alpha$}&{$\rm \beta$}\\
\noalign{\smallskip}
\hline 
\noalign{\smallskip}
\multicolumn{4}{c}{F555W =$\alpha$ + $\beta$ log P} \\
\hline
 0.0004 &   0.24    & -1.09 $\pm$ 0.23  & -2.99  $\pm$ 0.01 \\  
 0.004  &   0.25    & -1.45 $\pm$ 0.21  & -2.70  $\pm$ 0.02 \\
 0.008  &   0.25    & -1.48 $\pm$ 0.22  & -2.52  $\pm$ 0.02 \\
 0.01   &   0.26    & -1.33 $\pm$ 0.27  & -2.59  $\pm$ 0.02 \\
 0.02   &   0.25    & -1.32 $\pm$ 0.21  & -2.42  $\pm$ 0.02 \\
 0.02   &   0.26    & -1.29 $\pm$ 0.21  & -2.60  $\pm$ 0.02 \\
 0.02   &   0.28    & -1.67 $\pm$ 0.17  & -2.08  $\pm$ 0.01 \\
 0.02   &   0.31    & -1.54 $\pm$ 0.19  & -2.33  $\pm$ 0.02 \\
 0.03   &   0.275   & -1.34 $\pm$ 0.19  & -2.27  $\pm$ 0.02 \\
 0.03   &   0.31    & -1.56 $\pm$ 0.19  & -2.00  $\pm$ 0.01 \\
 0.03   &   0.335   & -1.49 $\pm$ 0.15  & -2.11  $\pm$ 0.01 \\
 0.04   &   0.25    & -1.22 $\pm$ 0.16  & -2.32  $\pm$ 0.01 \\
 0.04   &   0.29    & -1.46 $\pm$ 0.14  & -2.10  $\pm$ 0.01 \\
 0.04   &   0.33    & -1.40 $\pm$ 0.11  & -2.15  $\pm$ 0.01 \\                                                         
\hline 
\noalign{\smallskip}
\multicolumn{4}{c}{F606W =$\alpha$ + $\beta$ log P} \\
\hline
0.0004  &   0.24    & -1.25 $\pm$ 0.20  & -3.07  $\pm$ 0.01 \\  
 0.004  &   0.25    & -1.62 $\pm$ 0.19  & -2.79  $\pm$ 0.02 \\
 0.008  &   0.25    & -1.65 $\pm$ 0.20  & -2.63  $\pm$ 0.02 \\
 0.01   &   0.26    & -1.50 $\pm$ 0.24  & -2.70  $\pm$ 0.02 \\
 0.02   &   0.25    & -1.50 $\pm$ 0.19  & -2.54  $\pm$ 0.02 \\
 0.02   &   0.26    & -1.48 $\pm$ 0.19  & -2.71  $\pm$ 0.02 \\
 0.02   &   0.28    & -1.82 $\pm$ 0.16  & -2.23  $\pm$ 0.01 \\
 0.02   &   0.31    & -1.69 $\pm$ 0.17  & -2.46  $\pm$ 0.01 \\
 0.03   &   0.275   & -1.54 $\pm$ 0.17  & -2.41  $\pm$ 0.01 \\
 0.03   &   0.31    & -1.73 $\pm$ 0.17  & -2.16  $\pm$ 0.01 \\
 0.03   &   0.335   & -1.65 $\pm$ 0.14  & -2.28  $\pm$ 0.01 \\
 0.04   &   0.25    & -1.43 $\pm$ 0.15  & -2.46  $\pm$ 0.01 \\
 0.04   &   0.29    & -1.65 $\pm$ 0.13  & -2.25  $\pm$ 0.01 \\
 0.04   &   0.33    & -1.59 $\pm$ 0.10  & -2.30  $\pm$ 0.01 \\                                                         
\hline 
\noalign{\smallskip}
\multicolumn{4}{c}{F814W =$\alpha$ + $\beta$ log P} \\
\hline
 0.0004 &   0.24    & -1.68 $\pm$ 0.16  & -3.23  $\pm$ 0.01 \\  
 0.004  &   0.25    & -2.04 $\pm$ 0.15  & -2.96  $\pm$ 0.01 \\
 0.008  &   0.25    & -2.07 $\pm$ 0.15  & -2.83  $\pm$ 0.01 \\
 0.01   &   0.26    & -1.93 $\pm$ 0.19  & -2.91  $\pm$ 0.02 \\
 0.02   &   0.25    & -1.96 $\pm$ 0.15  & -2.77  $\pm$ 0.01 \\
 0.02   &   0.26    & -1.93 $\pm$ 0.15  & -2.91  $\pm$ 0.01 \\
 0.02   &   0.28    & -2.20 $\pm$ 0.12  & -2.50  $\pm$ 0.01 \\
 0.02   &   0.31    & -2.08 $\pm$ 0.14  & -2.71  $\pm$ 0.01 \\
 0.03   &   0.275   & -2.00 $\pm$ 0.14  & -2.65  $\pm$ 0.01 \\
 0.03   &   0.31    & -2.13 $\pm$ 0.13  & -2.46  $\pm$ 0.01 \\
 0.03   &   0.335   & -2.04 $\pm$ 0.11  & -2.58  $\pm$ 0.01 \\
 0.04   &   0.25    & -1.90 $\pm$ 0.12  & -2.72  $\pm$ 0.01 \\
 0.04   &   0.29    & -2.10 $\pm$ 0.11  & -2.53  $\pm$ 0.01 \\
 0.04   &   0.33    & -2.02 $\pm$ 0.08  & -2.57  $\pm$ 0.01 \\                                                         
\hline
\noalign{\smallskip}
\multicolumn{4}{c}{F160W =$\alpha$ + $\beta$ log P} \\
\hline
 0.0004 &   0.24    & -2.31 $\pm$ 0.09 & -3.454 $\pm$0.005 \\  
 0.004  &   0.25    & -2.63 $\pm$ 0.08 & -3.224 $\pm$0.007 \\
 0.008  &   0.25    & -2.62 $\pm$ 0.08 & -3.153 $\pm$0.006 \\
 0.01   &   0.26    & -2.50 $\pm$ 0.10 & -3.258 $\pm$0.009 \\
 0.02   &   0.25    & -2.53 $\pm$ 0.08 & -3.179 $\pm$0.007 \\
 0.02   &   0.26    & -2.52 $\pm$ 0.08 & -3.258 $\pm$0.007 \\
 0.02   &   0.28    & -2.67 $\pm$ 0.06 & -2.977 $\pm$0.004 \\
 0.02   &   0.31    & -2.56 $\pm$ 0.07 & -3.139 $\pm$0.006 \\
 0.03   &   0.275   & -2.57 $\pm$ 0.09 & -3.089 $\pm$0.007 \\
 0.03   &   0.31    & -2.60 $\pm$ 0.07 & -2.995 $\pm$0.005 \\
 0.03   &   0.335   & -2.48 $\pm$ 0.06 & -3.111 $\pm$0.005 \\
 0.04   &   0.25    & -2.48 $\pm$ 0.06 & -3.173 $\pm$0.005 \\
 0.04   &   0.29    & -2.64 $\pm$ 0.07 & -3.005 $\pm$0.005 \\
 0.04   &   0.33    & -2.54 $\pm$ 0.04 & -3.035 $\pm$0.003 \\                                                         
\hline
\hline
\end{tabular}
\label{tab1}
\end{table*}
\smallskip

\begin{table*}
\centering
\caption{As in Table \ref{tab1}, but for quadratic (left) and broken linear
  (right) PL--relations}
\begin{tabular}{llccc|cccc}
\hline
\hline
\noalign{\smallskip}
{$\rm Z$}&{$\rm Y$}&{$\rm \alpha$}&{$\rm \beta$}&{$\rm \gamma$}&{$\rm \alpha_{logP\le1}$}&{$\rm \beta_{logP\le1}$}&{$\rm \alpha_{logP>1}$}&{$\rm \beta_{logP>1}$}\\
\noalign{\smallskip}
\hline 
\noalign{\smallskip}
\multicolumn{9}{c}{F555W =$\alpha$ + $\beta$ log P  + $\gamma$ log P$^2$ ~~~~~~~~~~~~~~~~~~~~~~~~~~~~~~~~~~~~~~~~~          F555W =$\alpha$ + $\beta$ log P} \\
\hline
 0.0004 &   0.24  &-1.01$\pm$0.23 &  -3.17$\pm$ 0.07 &  0.09$\pm$0.03 &  -1.10$\pm$ 0.20 &  -2.96$\pm$0.03  &   -1.29$\pm$  0.28 &  -2.85$\pm$ 0.04\\     
 0.004  &   0.25  &-0.56$\pm$0.18 &  -4.65$\pm$ 0.09 &  0.93$\pm$0.04 &  -1.09$\pm$ 0.14 &  -3.20$\pm$0.03  &   -2.31$\pm$  0.26 &  -2.06$\pm$ 0.06\\  
 0.008  &   0.25  &-0.72$\pm$0.19 &  -4.19$\pm$ 0.09 &  0.79$\pm$0.04 &  -1.17$\pm$ 0.16 &  -2.95$\pm$0.04  &   -2.27$\pm$  0.26 &  -1.94$\pm$ 0.06\\  
 0.01   &   0.26  &-0.35$\pm$0.25 &  -4.71$\pm$ 0.14 &  1.03$\pm$0.07 &  -0.99$\pm$ 0.20 &  -3.05$\pm$0.05  &   -2.20$\pm$  0.33 &  -1.93$\pm$ 0.08\\  
 0.02   &   0.25  &-0.96$\pm$0.20 &  -3.28$\pm$ 0.10 &  0.45$\pm$0.05 &  -1.18$\pm$ 0.17 &  -2.63$\pm$0.03  &   -1.73$\pm$  0.27 &  -2.08$\pm$ 0.08\\  
 0.02   &   0.26  &-0.73$\pm$0.20 &  -3.96$\pm$ 0.10 &  0.72$\pm$0.05 &  -1.12$\pm$ 0.15 &  -2.86$\pm$0.03  &   -2.16$\pm$  0.29 &  -1.92$\pm$ 0.09\\  
 0.02   &   0.28  &-1.31$\pm$0.16 &  -2.88$\pm$ 0.06 &  0.38$\pm$0.03 &  -1.52$\pm$ 0.11 &  -2.29$\pm$0.02  &   -2.06$\pm$  0.23 &  -1.80$\pm$ 0.04\\  
 0.02   &   0.31  &-1.10$\pm$0.19 &  -3.27$\pm$ 0.10 &  0.45$\pm$0.05 &  -1.41$\pm$ 0.15 &  -2.49$\pm$0.04  &   -1.98$\pm$  0.24 &  -2.00$\pm$ 0.05\\  
 0.03   &   0.275 &-1.11$\pm$0.19 &  -2.81$\pm$ 0.09 &  0.28$\pm$0.04 &  -1.29$\pm$ 0.16 &  -2.33$\pm$0.03  &   -1.76$\pm$  0.22 &  -1.95$\pm$ 0.05\\  
 0.03   &   0.31  &-1.27$\pm$0.19 &  -2.63$\pm$ 0.09 &  0.30$\pm$0.04 &  -1.47$\pm$ 0.16 &  -2.12$\pm$0.04  &   -1.86$\pm$  0.23 &  -1.78$\pm$ 0.04\\  
 0.03   &   0.335 &-1.00$\pm$0.14 &  -3.13$\pm$ 0.07 &  0.46$\pm$0.03 &  -1.30$\pm$ 0.10 &  -2.36$\pm$0.03  &   -1.86$\pm$  0.18 &  -1.84$\pm$ 0.03\\  
 0.04   &   0.25  &-1.13$\pm$0.16 &  -2.56$\pm$ 0.07 &  0.14$\pm$0.04 &  -1.19$\pm$ 0.14 &  -2.35$\pm$0.02  &   -1.53$\pm$  0.20 &  -2.08$\pm$ 0.07\\  
 0.04   &   0.29  &-1.26$\pm$0.14 &  -2.60$\pm$ 0.06 &  0.26$\pm$0.03 &  -1.40$\pm$ 0.12 &  -2.20$\pm$0.02  &   -1.86$\pm$  0.17 &  -1.80$\pm$ 0.04\\  
 0.04   &   0.33  &-1.09$\pm$0.10 &  -2.82$\pm$ 0.05 &  0.32$\pm$0.02 &  -1.29$\pm$ 0.08 &  -2.29$\pm$0.02  &   -1.72$\pm$  0.13 &  -1.92$\pm$ 0.03\\  
\hline 
\noalign{\smallskip}
\multicolumn{9}{c}{F606W =$\alpha$ + $\beta$ log P  + $\gamma$ log  P$^2$ ~~~~~~~~~~~~~~~~~~~~~~~~~~~~~~~~~~~~~~~~~         F606W =$\alpha$ + $\beta$ log P} \\
\hline
 0.0004 &   0.24    &-1.19$\pm$ 0.20 &  -3.20$\pm$ 0.07 &  0.06$\pm$ 0.03 &  -1.27$\pm$ 0.18 &  -3.03$\pm$ 0.03 &   -1.42$\pm$  0.25 &  -2.95$\pm$ 0.04\\  
 0.004  &   0.25    &-0.82$\pm$ 0.17 &  -4.53$\pm$ 0.08 &  0.83$\pm$ 0.04 &  -1.29$\pm$ 0.13 &  -3.24$\pm$ 0.03 &   -2.38$\pm$  0.24 &  -2.22$\pm$ 0.05\\  
 0.008  &   0.25    &-0.96$\pm$ 0.17 &  -4.12$\pm$ 0.08 &  0.71$\pm$ 0.04 &  -1.37$\pm$ 0.14 &  -3.01$\pm$ 0.03 &   -2.35$\pm$  0.24 &  -2.11$\pm$ 0.05\\  
 0.01   &   0.26    &-0.62$\pm$ 0.22 &  -4.61$\pm$ 0.13 &  0.93$\pm$ 0.06 &  -1.20$\pm$ 0.18 &  -3.12$\pm$ 0.05 &   -2.29$\pm$  0.30 &  -2.10$\pm$ 0.07\\  
 0.02   &   0.25    &-1.18$\pm$ 0.18 &  -3.33$\pm$ 0.09 &  0.41$\pm$ 0.05 &  -1.38$\pm$ 0.15 &  -2.74$\pm$ 0.03 &   -1.89$\pm$  0.25 &  -2.24$\pm$ 0.07\\  
 0.02   &   0.26    &-0.98$\pm$ 0.18 &  -3.94$\pm$ 0.09 &  0.65$\pm$ 0.05 &  -1.33$\pm$ 0.14 &  -2.94$\pm$ 0.03 &   -2.26$\pm$  0.26 &  -2.09$\pm$ 0.08\\  
 0.02   &   0.28    &-1.49$\pm$ 0.15 &  -2.96$\pm$ 0.06 &  0.35$\pm$ 0.03 &  -1.69$\pm$ 0.10 &  -2.42$\pm$ 0.02 &   -2.18$\pm$  0.21 &  -1.97$\pm$ 0.04\\  
 0.02   &   0.31    &-1.29$\pm$ 0.17 &  -3.32$\pm$ 0.09 &  0.41$\pm$ 0.04 &  -1.58$\pm$ 0.13 &  -2.61$\pm$ 0.03 &   -2.10$\pm$  0.22 &  -2.16$\pm$ 0.05\\  
 0.03   &   0.275   &-1.32$\pm$ 0.17 &  -2.92$\pm$ 0.08 &  0.26$\pm$ 0.04 &  -1.49$\pm$ 0.15 &  -2.47$\pm$ 0.03 &   -1.93$\pm$  0.21 &  -2.11$\pm$ 0.05\\  
 0.03   &   0.31    &-1.45$\pm$ 0.17 &  -2.76$\pm$ 0.08 &  0.28$\pm$ 0.04 &  -1.64$\pm$ 0.14 &  -2.23$\pm$ 0.03 &   -2.01$\pm$  0.20 &  -1.96$\pm$ 0.04\\  
 0.03   &   0.335   &-1.19$\pm$ 0.12 &  -3.22$\pm$ 0.06 &  0.43$\pm$ 0.03 &  -1.47$\pm$ 0.09 &  -2.51$\pm$ 0.02 &   -2.00$\pm$  0.17 &  -2.03$\pm$ 0.03\\  
 0.04   &   0.25    &-1.34$\pm$ 0.15 &  -2.70$\pm$ 0.07 &  0.13$\pm$ 0.04 &  -1.40$\pm$ 0.13 &  -2.50$\pm$ 0.02 &   -1.71$\pm$  0.18 &  -2.23$\pm$ 0.06\\  
 0.04   &   0.29    &-1.46$\pm$ 0.13 &  -2.73$\pm$ 0.05 &  0.24$\pm$ 0.03 &  -1.59$\pm$ 0.11 &  -2.34$\pm$ 0.02 &   -2.04$\pm$  0.16 &  -1.97$\pm$ 0.04\\  
 0.04   &   0.33    &-1.30$\pm$ 0.09 &  -2.93$\pm$ 0.04 &  0.29$\pm$ 0.02 &  -1.49$\pm$ 0.07 &  -2.43$\pm$ 0.02 &   -1.88$\pm$  0.12 &  -2.09$\pm$ 0.02\\  
\hline 
\noalign{\smallskip}
\multicolumn{9}{c}{F814W =$\alpha$ + $\beta$ log P  + $\gamma$ log  P$^2$ ~~~~~~~~~~~~~~~~~~~~~~~~~~~~~~~~~~~~~~~~~         F814W =$\alpha$ + $\beta$ log P} \\
\hline
 0.0004 &   0.24    &-1.66$\pm$ 0.16 &  -3.28$\pm$0.05 &  0.03$\pm$  0.02 &  -1.70$\pm$ 0.14   &  -3.19$\pm$0.02 &    -1.79$\pm$ 0.19   & -3.15$\pm$ 0.03\\    
 0.004  &   0.25    &-1.42$\pm$ 0.13 &  -4.31$\pm$0.06 &  0.65$\pm$  0.03 &  -1.79$\pm$ 0.10   &  -3.31$\pm$0.02 &    -2.63$\pm$ 0.18   & -2.52$\pm$ 0.04\\    
 0.008  &   0.25    &-1.52$\pm$ 0.14 &  -4.02$\pm$0.06 &  0.57$\pm$  0.03 &  -1.85$\pm$ 0.11   &  -3.13$\pm$0.02 &    -2.63$\pm$ 0.19   & -2.41$\pm$ 0.04\\    
 0.01   &   0.26    &-1.22$\pm$ 0.18 &  -4.44$\pm$0.10 &  0.74$\pm$  0.05 &  -1.69$\pm$ 0.14   &  -3.24$\pm$0.04 &    -2.56$\pm$ 0.24   & -2.43$\pm$ 0.06\\    
 0.02   &   0.25    &-1.69$\pm$ 0.15 &  -3.41$\pm$0.07 &  0.34$\pm$  0.04 &  -1.85$\pm$ 0.12   &  -2.93$\pm$0.02 &    -2.26$\pm$ 0.20   & -2.52$\pm$ 0.06\\    
 0.02   &   0.26    &-1.52$\pm$ 0.14 &  -3.91$\pm$0.08 &  0.53$\pm$  0.04 &  -1.81$\pm$ 0.11   &  -3.10$\pm$0.02 &    -2.55$\pm$ 0.21   & -2.41$\pm$ 0.06\\    
 0.02   &   0.28    &-1.94$\pm$ 0.12 &  -3.09$\pm$0.04 &  0.28$\pm$  0.02 &  -2.09$\pm$ 0.08   &  -2.66$\pm$0.02 &    -2.50$\pm$ 0.16   & -2.29$\pm$ 0.03\\    
 0.02   &   0.31    &-1.75$\pm$ 0.13 &  -3.42$\pm$0.07 &  0.34$\pm$  0.03 &  -1.98$\pm$ 0.10   &  -2.84$\pm$0.02 &    -2.42$\pm$ 0.17   & -2.46$\pm$ 0.04\\    
 0.03   &   0.275   &-1.82$\pm$ 0.14 &  -3.07$\pm$0.07 &  0.21$\pm$  0.03 &  -1.96$\pm$ 0.12   &  -2.70$\pm$0.03 &    -2.32$\pm$ 0.17   & -2.41$\pm$ 0.04\\    
 0.03   &   0.31    &-1.92$\pm$ 0.13 &  -2.90$\pm$0.06 &  0.21$\pm$  0.03 &  -2.06$\pm$ 0.11   &  -2.56$\pm$0.03 &    -2.33$\pm$ 0.16   & -2.31$\pm$ 0.03\\    
 0.03   &   0.335   &-1.68$\pm$ 0.10 &  -3.30$\pm$0.05 &  0.33$\pm$  0.02 &  -1.89$\pm$ 0.07   &  -2.77$\pm$0.02 &    -2.30$\pm$ 0.13   & -2.39$\pm$ 0.03\\    
 0.04   &   0.25    &-1.82$\pm$ 0.12 &  -2.91$\pm$0.05 &  0.11$\pm$  0.03 &  -1.87$\pm$ 0.10   &  -2.75$\pm$0.02 &    -2.11$\pm$ 0.15   & -2.55$\pm$ 0.05\\    
 0.04   &   0.29    &-1.95$\pm$ 0.11 &  -2.89$\pm$0.04 &  0.19$\pm$  0.02 &  -2.05$\pm$ 0.09   &  -2.60$\pm$0.02 &    -2.39$\pm$ 0.13   & -2.31$\pm$ 0.03\\    
 0.04   &   0.33    &-1.80$\pm$ 0.07 &  -3.04$\pm$0.03 &  0.22$\pm$  0.02 &  -1.94$\pm$ 0.06   &  -2.67$\pm$0.01 &    -2.25$\pm$ 0.09   & -2.40$\pm$ 0.02\\    
\hline 
\noalign{\smallskip}
\multicolumn{9}{c}{F160W =$\alpha$ + $\beta$ log P  + $\gamma$ log  P$^2$  ~~~~~~~~~~~~~~~~~~~~~~~~~~~~~~~~~~~~~~~~~        F160W =$\alpha$ + $\beta$ log P} \\
\hline
 0.0004 &   0.24    &-2.30$\pm$ 0.09 &  -3.47$\pm$ 0.03 &  0.01$\pm$  0.01 &  -2.33$\pm$0.08  & -3.43$\pm$ 0.01 &    -2.37$\pm$ 0.10  & -3.42$\pm$0.02\\    
 0.004  &   0.25    &-2.31$\pm$ 0.07 &  -3.92$\pm$ 0.03 &  0.33$\pm$  0.02 &  -2.50$\pm$0.05  & -3.40$\pm$ 0.01 &    -2.93$\pm$ 0.10  & -3.00$\pm$0.02\\    
 0.008  &   0.25    &-2.34$\pm$ 0.07 &  -3.76$\pm$ 0.03 &  0.29$\pm$  0.02 &  -2.51$\pm$0.06  & -3.31$\pm$ 0.01 &    -2.91$\pm$ 0.10  & -2.94$\pm$0.02\\    
 0.01   &   0.26    &-2.13$\pm$ 0.09 &  -4.06$\pm$ 0.05 &  0.39$\pm$  0.03 &  -2.37$\pm$0.08  & -3.43$\pm$ 0.02 &    -2.83$\pm$ 0.13  & -3.01$\pm$0.03\\    
 0.02   &   0.25    &-2.40$\pm$ 0.08 &  -3.50$\pm$ 0.04 &  0.17$\pm$  0.02 &  -2.48$\pm$0.06  & -3.26$\pm$ 0.01 &    -2.69$\pm$ 0.11  & -3.05$\pm$0.03\\    
 0.02   &   0.26    &-2.30$\pm$ 0.08 &  -3.78$\pm$ 0.04 &  0.28$\pm$  0.02 &  -2.45$\pm$0.06  & -3.35$\pm$ 0.01 &    -2.85$\pm$ 0.11  & -3.00$\pm$0.03\\    
 0.02   &   0.28    &-2.54$\pm$ 0.06 &  -3.26$\pm$ 0.02 &  0.13$\pm$  0.01 &  -2.62$\pm$0.04  & -3.05$\pm$ 0.01 &    {\bf-2.81$\pm$ 0.08} & {\bf -2.88$\pm$0.01}\\    
 0.02   &   0.31    &-2.39$\pm$ 0.07 &  -3.50$\pm$ 0.04 &  0.18$\pm$  0.02 &  -2.51$\pm$0.05  & -3.20$\pm$ 0.01 &    -2.73$\pm$ 0.09  & -3.01$\pm$0.02\\    
 0.03   &   0.275   &-2.48$\pm$ 0.09 &  -3.30$\pm$ 0.04 &  0.11$\pm$  0.02 &  -2.55$\pm$0.08  & -3.11$\pm$ 0.02 &    -2.74$\pm$ 0.10  & -2.96$\pm$0.02\\    
 0.03   &   0.31    &-2.51$\pm$ 0.07 &  -3.18$\pm$ 0.03 &  0.09$\pm$  0.01 &  -2.57$\pm$0.06  & -3.03$\pm$ 0.01 &    -2.68$\pm$ 0.08  & -2.93$\pm$0.02\\    
 0.03   &   0.335   &-2.31$\pm$ 0.05 &  -3.45$\pm$ 0.03 &  0.16$\pm$  0.01 &  -2.41$\pm$0.04  & -3.20$\pm$ 0.01 &    -2.60$\pm$ 0.07  & -3.02$\pm$0.01\\    
 0.04   &   0.25    &-2.45$\pm$ 0.06 &  -3.24$\pm$ 0.03 &  0.04$\pm$  0.02 &  -2.47$\pm$0.06  & -3.18$\pm$ 0.01 &    -2.57$\pm$ 0.08  & -3.10$\pm$0.03\\    
 0.04   &   0.29    &-2.57$\pm$ 0.07 &  -3.17$\pm$ 0.03 &  0.09$\pm$  0.01 &  -2.62$\pm$0.06  & -3.03$\pm$ 0.01 &    -2.79$\pm$ 0.08  & -2.90$\pm$0.02\\    
 0.04   &   0.33    &-2.44$\pm$ 0.04 &  -3.25$\pm$ 0.02 &  0.10$\pm$  0.01 &  -2.50$\pm$0.03  & -3.08$\pm$ 0.01 &    -2.64$\pm$ 0.05  & -2.96$\pm$0.01\\    
\hline
\hline
\end{tabular}
\label{tab2}
\end{table*}
\smallskip


\begin{table*}
\centering
\caption{Wesenheit--relation for different
  chemical compositions.}
\begin{tabular}{lccc}
\hline
\hline
\noalign{\smallskip}
{$\rm Z$}&{$\rm Y$}&{$\rm \alpha$}&{$\rm \beta$}\\
\noalign{\smallskip}
\hline
\noalign{\smallskip}
\multicolumn{4}{c}{F555W-2.42(F555W-F814W)}=$\alpha$ + $\beta$ log P \\
\hline
 0.0004 &   0.24    & -2.53$\pm$     0.06 & -3.563$\pm$    0.004\\
 0.004  &   0.25    & -2.87$\pm$     0.06 & -3.329$\pm$    0.006\\
 0.008  &   0.25    & -2.89$\pm$     0.06 & -3.260$\pm$    0.005\\
 0.01   &   0.26    & -2.79$\pm$     0.08 & -3.359$\pm$    0.008\\
 0.02   &   0.25    & -2.87$\pm$     0.07 & -3.275$\pm$    0.006\\
 0.02   &   0.26    & -2.84$\pm$     0.07 & -3.352$\pm$    0.006\\
 0.02   &   0.28    & -2.97$\pm$     0.05 & -3.096$\pm$    0.003\\
 0.02   &   0.31    & -2.86$\pm$     0.06 & -3.254$\pm$    0.005\\
 0.03   &   0.275   & -2.93$\pm$     0.08 & -3.196$\pm$    0.006\\
 0.03   &   0.31    & -2.94$\pm$     0.06 & -3.125$\pm$    0.004\\
 0.03   &   0.335   & -2.81$\pm$     0.05 & -3.240$\pm$    0.004\\
 0.04   &   0.25    & -2.85$\pm$     0.05 & -3.291$\pm$    0.004\\
 0.04   &   0.29    & -3.01$\pm$     0.06 & -3.128$\pm$    0.005\\
 0.04   &   0.33    & -2.91$\pm$     0.03 & -3.155$\pm$    0.003\\
\hline
\noalign{\smallskip}
\multicolumn{4}{c}{F606W-2.82(F606W-F814W)}=$\alpha$ + $\beta$ log P \\
\hline
 0.0004 &   0.24    & -2.47$\pm$     0.07 & -3.518$\pm$    0.004\\
 0.004  &   0.25    & -2.81$\pm$     0.07 & -3.273$\pm$    0.006\\
 0.008  &   0.25    & -2.83$\pm$     0.08 & -3.190$\pm$    0.006\\
 0.01   &   0.26    & -2.72$\pm$     0.10 & -3.285$\pm$    0.009\\
 0.02   &   0.25    & -2.78$\pm$     0.08 & -3.186$\pm$    0.007\\
 0.02   &   0.26    & -2.76$\pm$     0.08 & -3.273$\pm$    0.007\\
 0.02   &   0.28    & -2.91$\pm$     0.06 & -2.994$\pm$    0.004\\
 0.02   &   0.31    & -2.79$\pm$     0.07 & -3.158$\pm$    0.006\\
 0.03   &   0.275   & -2.83$\pm$     0.09 & -3.098$\pm$    0.007\\
 0.03   &   0.31    & -2.86$\pm$     0.07 & -3.013$\pm$    0.005\\
 0.03   &   0.335   & -2.74$\pm$     0.06 & -3.127$\pm$    0.005\\
 0.04   &   0.25    & -2.75$\pm$     0.06 & -3.188$\pm$    0.005\\
 0.04   &   0.29    & -2.91$\pm$     0.07 & -3.022$\pm$    0.005\\
 0.04   &   0.33    & -2.81$\pm$     0.04 & -3.050$\pm$    0.003\\
\hline
\noalign{\smallskip}
\multicolumn{4}{c}{F160W-0.41(V-I)}=$\alpha$ + $\beta$ log P \\
\hline
 0.0004 &   0.24      & -2.54$\pm$     0.06 & -3.540$\pm$    0.004\\
 0.004  &   0.25      & -2.85$\pm$     0.05 & -3.320$\pm$    0.006\\
 0.008  &   0.25      & -2.84$\pm$     0.06 & -3.271$\pm$    0.004\\
 0.01   &   0.26      & -2.73$\pm$     0.07 & -3.382$\pm$    0.007\\
 0.02   &   0.25      & -2.78$\pm$     0.06 & -3.315$\pm$    0.005\\
 0.02   &   0.26      & -2.76$\pm$     0.06 & -3.376$\pm$    0.005\\
 0.02   &   0.28      & -2.87$\pm$     0.04 & -3.138$\pm$    0.003\\
 0.02   &   0.31      & -2.76$\pm$     0.05 & -3.285$\pm$    0.004\\
 0.03   &   0.275     & -2.81$\pm$     0.07 & -3.243$\pm$    0.006\\
 0.03   &   0.31      & -2.81$\pm$     0.05 & -3.182$\pm$    0.004\\
 0.03   &   0.335     & -2.69$\pm$     0.04 & -3.299$\pm$    0.003\\
 0.04   &   0.25      & -2.74$\pm$     0.04 & -3.340$\pm$    0.004\\
 0.04   &   0.29      & -2.89$\pm$     0.05 & -3.181$\pm$    0.004\\
 0.04   &   0.33      & -2.78$\pm$     0.03 & -3.208$\pm$    0.002\\
\hline
\end{tabular}
\label{tab3}
\end{table*}

\begin{table*}
\centering
\caption{Global properties of the SNIa host galaxies observed with
  WFC3@HST (see text for details).}
\begin{tabular}{lcccc|cccc}
\hline
\hline
{$\rm ID$}&{$\rm N_{Cep}$(tot/{\it good})}&{$\rm <12+log[O/H]>$} &$\rm m_{v,i}^0 +5a_v$&{$\rm \mu_0$(Riess)}&{$\rm
  \mu_0$(this paper)} & {$\rm A_{V}$}& {$\rm A_{I}$}& {$\rm A_{F160W}$} \\
\hline
\noalign{\smallskip}
NGC 4258 &167/117&8.89$^a$ $\pm$0.09& & 29.29& 29.345$\pm$0.004$^b$(0.34) & 0.044& 0.024&0.009 \\
\hline            
\hline
NGC 4536 &69/49 &8.77$^a$ $\pm$0.14 & 15.147 &30.91$\pm$0.07& 30.92$\pm$0.05$^b$(0.32)& 0.050 & 0.027 &0.010\\
NGC 4639 &37/25 &8.98$^a$ $\pm$0.15 & 16.040 &31.67$\pm$0.08& 31.60$\pm$0.08$^b$(0.40)& 0.071 & 0.039 &0.015\\
NGC 3370 &101/69&8.81$^a$ $\pm$0.14 & 16.545 &32.13$\pm$0.07& 32.04$\pm$0.03$^b$(0.32)& 0.084 & 0.046 &0.018\\
NGC 3982 &38/22 &8.77$^a$ $\pm$0.24 & 15.953 &31.70$\pm$0.08& 31.79$\pm$0.06$^b$(0.32)& 0.039 & 0.021 &0.008\\
NGC 3021 &30/22 &8.89$^a$ $\pm$0.18 & 16.699 &32.27$\pm$0.08& 32.12$\pm$0.08$^b$(0.39)& 0.037 & 0.020 &0.008\\
NGC 1309 &56/30 &8.90$^a$ $\pm$0.14 & 16.768 &32.59$\pm$0.09& 32.40$\pm$0.06$^b$(0.38)& 0.109 & 0.060 &0.023\\
NGC 5584 &103/78&8.84$^a$ $\pm$0.12 & 16.274 &31.72$\pm$0.07& 31.71$\pm$0.04$^b$(0.32)& 0.107 & 0.058 &0.022\\
NGC 4038 &46/32 &9.01$^a$ $\pm$0.09 & 15.901 &31.66$\pm$0.08& 31.55$\pm$0.06$^b$(0.36)& 0.127 & 0.070 &0.008\\
\hline
\hline
\end{tabular}
\label{tab4}
\\
~~
$^a$ Metallicity content from \citet{riess11a};\\
$^b$These errors have been computed using the standard deviation
$\sigma$ (in parenthesis) of $< \mu_0 >$ divided by $\sqrt{N_{Cep}({\it tot}})$, see Fig.~\ref{fig6}.
\end{table*}
\smallskip
\smallskip

\bibliographystyle{mn2e}


\end{document}